\documentclass[pra, notitlepage, showpacs, floatfix, reprint, citeautoscript]{revtex4-1}

\usepackage[T1]{fontenc}
\usepackage[latin9]{inputenc}

\usepackage{lipsum}
\usepackage{amsmath}
\usepackage{amssymb}
\usepackage{bbm}
\usepackage{braket}
\usepackage{xcolor}
\usepackage{pifont}
\usepackage[mathscr]{euscript}
\usepackage[shortlabels]{enumitem}

\allowdisplaybreaks

\usepackage{graphicx}
\usepackage[colorlinks=true]{hyperref}  
\hypersetup{
    bookmarks=true,         
    unicode=false,          
    pdftoolbar=true,        
    pdfmenubar=true,        
    pdffitwindow=false,     
    pdfstartview={FitH},    
    pdftitle={Unquantized thermal Hall effect in quantum spin liquids with spinon Fermi surfaces},    
    pdfauthor={Rhine Samajdar},     
    pdfsubject={},   
    pdfcreator={},   
    pdfproducer={}, 
    pdfkeywords={} {} {}, 
    pdfnewwindow=true,      
    colorlinks=true,       
    linkcolor=blue, 
    citecolor=blue,        
    filecolor=magenta,      
    urlcolor=blue           
}

\newcommand{\equref}[1]{Eq.~(\ref{#1})}

\newcommand{\pdagger}{{\phantom{\dagger}}}

\renewcommand{\approx}{\simeq}

\newcommand{\mc}[1]{\mathcal{#1}}
\renewcommand{\vec}[1]{\boldsymbol{#1}}

\newcommand{\ie}{i.e.~}

\definecolor{wrongultramarine}{rgb}{1,0.5,0}

\begin{document}

\title{Unquantized thermal Hall effect in quantum spin liquids with spinon Fermi surfaces}

\author{Yanting Teng}
\affiliation{Department of Physics, Harvard University, Cambridge MA 02138, USA}

\author{Yunchao Zhang}
\affiliation{Department of Physics, Harvard University, Cambridge MA 02138, USA}

\author{Rhine Samajdar}
\affiliation{Department of Physics, Harvard University, Cambridge MA 02138, USA}

\author{Mathias S. Scheurer}
\affiliation{Department of Physics, Harvard University, Cambridge MA 02138, USA}

\author{Subir Sachdev}
\affiliation{Department of Physics, Harvard University, Cambridge MA 02138, USA}

\begin{abstract}
Recent theoretical studies have found quantum spin liquid states with spinon Fermi surfaces upon the application of a magnetic field on a gapped state with topological order. We investigate the thermal Hall conductivity across this transition, describing how the quantized thermal Hall conductivity of the gapped state changes to an unquantized thermal Hall conductivity in the gapless spinon Fermi surface state. We consider two cases, both of potential experimental interest: the state with non-Abelian Ising topological order on the honeycomb lattice, and the state with Abelian chiral spin liquid topological order on the triangular lattice.
\end{abstract}

\maketitle

\section{Introduction}

Quantum spin liquids (QSLs) are highly correlated systems of mutually interacting spins, in which zero-point quantum fluctuations are so strong as to prevent symmetry-breaking magnetic ordering down to the lowest temperatures \cite{anderson1973resonating,balents2010spin,knolle2019field,broholm2020quantum}. More generally speaking, QSLs are best defined as phases of matter realizing ground states with long-range many-body entanglement, or massive quantum superposition \cite{savary2016quantum}. The exotic properties of these highly entangled states can often be better understood in terms of new (possibly nonlocal) degrees of freedom rather than the constituent spins themselves \cite{zhou2017quantum}. Indeed, quite generically, QSLs are characterized by ``fractionalized'' excitations such as  charge-neutral spinons. These spinons, which are accompanied by emergent gauge fields, may or may not possess an energy gap, and can obey either Fermi or Bose statistics \cite{read1991large,senthil2000z, senthil2001fractionalization}.

First proposed in the 1970s, QSLs eluded experimental discovery in magnetic compounds for nearly half a century and even today, undisputed material candidates are few and far in between \cite{Lee1306, norman2016}. On the theoretical side, however, models of these enigmatic phases are plentiful. The prototypical example of a system with an \textit{exact} spin-liquid ground state is the Kitaev model \cite{kitaev2006anyons}. When placed in a magnetic field, this model hosts a gapped phase with topological order, supporting Majorana fermions and non-Abelian Ising anyons \cite{nayak2008non}, which may be relevant for quantum computation \cite{stern2013topological}. Despite the seemingly contrived form of the bond-directional interactions in the Kitaev model, variants thereof can actually be realized in some spin-orbit entangled $j =1/2$ Mott insulators 
\cite{khaliullin2005orbital, jackeli2009mott, chaloupka2010kitaev}. Among these so-called ``Kitaev materials'' \cite{witczak2014correlated, rau2016spin,trebst2017kitaev,winter2017models,hermanns2018physics} are layered iridates such as Na$_2$IrO$_3$ \cite{singhAFM, chun2015direct} and La$_2$IrO$_3$ \cite{singh2012relevance}, where the iridium atoms form the sites of a honeycomb lattice.

Another promising material in this family, which has attracted much attention recently, is $\alpha$-RuCl$_3$; here, the Ru$^{3+}$ ions act as effective localized moments. The ground state of $\alpha$-RuCl$_3$ is known to be magnetically ordered \cite{fletcher1967x, kobayashi1992moessbauer} in the absence of a Zeeman field, with a zigzag antiferromagnetic pattern \cite{sears2015magnetic, johnson2015monoclinic, cao2016low}. While the system orders, in zero field, at about 7~$\mathrm{K}$ \cite{zigzag2013,rau2014generic}, the Kitaev exchange interaction is estimated to be $\sim$ 50--90~$\mathrm{K}$. 
This wide separation of scales has been interpreted as evidence for proximity to Kitaev's QSL state. However, searching for fingerprints of charge-neutral quasiparticles that could unambiguously identify this state is challenging with the familiar techniques that rely on electrical transport. 
In this regard, a powerful probe of unconventional excitations in insulators is the thermal Hall effect, also known as the Righi-Leduc effect. For instance, recent measurements of a giant thermal Hall conductivity in several undoped cuprate superconductors \cite{grissonnanche2019giant, grissonnanche2020phonons} have offered new insights \cite{prb2019,han2019consideration,2019enhanced,li2019thermal,chen2019enhanced,guo2020gauge} into their underlying electronic phases. 

The thermal Hall effect is especially of relevance to the Kitaev materials because even if the charge degrees of freedom are frozen out, heat transport \cite{leahy2017anomalous} can still be facilitated through charge-neutral modes. In $\alpha$-RuCl$_3$, upon applying a Zeeman field, the intrinsic zigzag order melts \cite{banerjee2017neutron,lampen2018field,janvsa2018observation,banerjee2018excitations}, driving the system into a paramagnetic phase. If the field induces the aforementioned topologically ordered phase, which has a chiral Majorana fermion edge state, one would expect a half-quantized [in units of $(\pi/6) k_B^2/\hbar$] thermal Hall response \cite{banerjee2018observation} as $T\rightarrow 0$. Claims of such observations \cite{kasahara2018majorana}, suggesting a non-Abelian Ising anyon phase, have sparked extensive investigation, both experimentally \cite{kasahara2018unusual,balz2019finite,lefrancois2019thermal, yokoi2020half, yamashita2020} and theoretically \cite{ye2018quantization,vinkler2018approximately,cookmeyer2018spin,egmoon2020}. Curiously enough, a finite but unquantized thermal Hall
conductivity was also measured in $\alpha$-RuCl$_3$, over a broad range of temperatures and magnetic fields \cite{kasahara2018unusual, hentrich2018large}. This points towards a scenario where the effect of the field yields an additional $\mathrm{U}(1)$ QSL phase \cite{gao2019topological}. Indeed a plethora of numerical studies \cite{zhu2018robust, liang2018intermediate, gohlke2018dynamical,nasu2018successive,hickey2019emergence,ronquillo2019signatures,patel2019magnetic} indicate the presence of an intermediate gapless phase with spinon Fermi surfaces (SFS), between the gapped topological order and the trivial polarized phase at very strong fields. 

Motivated by these diverse observations, we examine the thermal Hall response in the Kitaev model for a wide variety of field strengths and orientations using a parton mean-field theory \cite{burnell20112,schaffer2012,okamoto2013}. One of our goals will be to understand the half-quantized conductivity---and its stability---as a function of an applied magnetic field. The quantization ceases to hold as the system undergoes a phase transition \cite{sachdev_2011} to a field-induced U($1$) spin liquid. We systematically investigate the thermal Hall signatures of this gapless phase, including its temperature and field dependence, and show that it is consistent with the behavior seen in experiments. In particular, we demonstrate that an unquantized response can be obtained from simply the pure Kitaev model, coupled to a field, without requiring any of the auxiliary Dzyaloshinskii-Moriya interactions assumed in Ref.~\cite{gao2019thermal}.

Interestingly, similar gapless QSLs with Fermi surfaces of neutral emergent excitations can also appear in systems lacking spin-orbit coupling. One of the most commonly studied examples of this type is the Heisenberg model on a triangular lattice. The physical importance of this simple model is paramount as sundry QSL candidates fall in the category of layered spin-$1/2$ triangular-lattice magnets, like the organic salts~\cite{shimizu2003, kurosaki2005, yamashita2008, yamashita2009, yamashita2010} and the transition metal dichalcogenides~\cite{Law6996, yu2017, ribak2017, klanjsek2017}.
In all these materials, which belong to the family of weak Mott insulators with strong charge fluctuations~\cite{misguich1999, motrunich2005variational, sheng2009spin, block2011, he2018}, transport measurements hint at the existence of extensive mobile gapless spin excitations. While, conventionally, many of the Hamiltonians used to describe these compounds have included  ``ring-exchange'' couplings involving multiple spins, replacing these with competing two-spin interactions between different neighboring sites leads to equally rich physics. In fact, such competition between neighboring couplings has proved to be an essential ingredient in understanding potential QSL states in the triangular-lattice delafossites \cite{kadowaki1990, kimura2006, ye2007, seki2008, daidai2020} and rare-earth compounds~\cite{li2015, shen2016evidence,paddison2017,li2017,paddison2017, zhu2018, zhang2018, shen2018}. Guided by recent numerical work identifying a gapless chiral spin liquid (CSL) phase on the triangular lattice \cite{gong2019chiral}, we analyze the thermal Hall coefficient in a Heisenberg antiferromagnet with competing exchange terms up to third-nearest neighbors. Furthermore, we compare and contrast this conductivity to that calculated for $\alpha$-RuCl$_3$ earlier. Our work highlights how two seemingly disparate systems---the Kitaev and Heisenberg models---exhibit parallel unquantized thermal Hall responses, and underscores the generality of the same.

\section{Kitaev honeycomb model}
\label{sec:majorana}

\begin{figure}[tb]
\includegraphics[width=\linewidth]{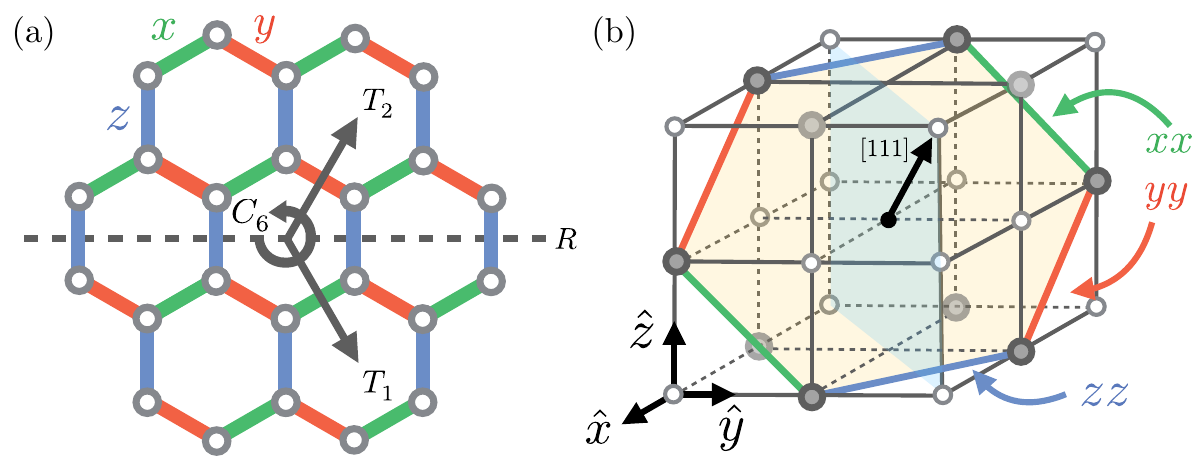}
\caption{\label{fig:symmetry}(a) The Kitaev model on the honeycomb lattice. Each hexagon has three types of links, labeled by $x$, $y$, or $z$; the interactions between nearest-neighboring spins are bond-dependent, as prescribed by Eq.~\eqref{eq:kitaev}. The operations  $R$ and $C_6$ designate reflection across the $x=y$ plane, and  sixfold $c$-axis rotation, respectively. (b) A hexagonal plaquette embedded in the three-dimensional cubic lattice. The $c$ axis is oriented along the $[111]$ direction. The lattice plane in (a), $c=0$, is shaded in yellow, while the blue shading marks the $ac$ plane (containing the magnetic field directions studied here). }
\end{figure}

Kitaev's eponymous model, introduced in Ref.~\cite{kitaev2006anyons}, is comprised of  $S = 1/2$ spins arranged on a honeycomb lattice, with the Hamiltonian:
\begin{equation}
\label{eq:kitaev}
H^{}_{\textsc{k}}= K^{}_{x} \sum_{x\,\,\mathrm{links}} S^x_j S^x_j +K^{}_{y} \sum_{y\,\,\mathrm{links}} S^y_j S^y_j +K^{}_{z} \sum_{z\,\,\mathrm{links}} S^z_j S^z_j,
\end{equation}
where $\vec{S}_j$\,$=$\,$(S_j^x,S_j^y,S_j^z)$ represents the spin operator at site $j$.
The spin and orbital degrees of freedom are locally entangled as the interactions between nearest neighbors depend on the type of the link. There are three nonequivalent bond directions: the $z$ links are the vertical bonds of the lattice, whereas the bonds angled at $\pm \pi/3$ from the vertical constitute the $x$ and $y$ links (see Fig.~\ref{fig:symmetry}).  A remarkable feature of this model is that it is exactly solvable and hosts different QSL \cite{savary2016quantum, zhou2017quantum, broholm2020quantum} ground states.  It has a gapped topological phase (known as the $A$ phase), which is equivalent to the toric code \cite{kitaev2003fault} and supports Abelian anyons. It also has a gapless phase (the $B$ phase), which morphs to a gapped non-Abelian topological phase, realizing Ising topological order (ITO), under a time-reversal symmetry-breaking perturbation. The fractionalized excitations in this gapped QSL are Majorana fermions and Ising anyons. 

Before studying the Kitaev model in a magnetic field, it is instructive to first consider how the zero-field model can be solved by writing the spins in terms of Majorana fermions. This will also help us to draw a distinction with the fermionization procedure employed later in Sec.~\ref{sec:parton}. We first define, for each site $j$, four Majorana fermions  $\{\chi^\mu\}$ such that:
\begin{alignat}{1}\label{majorana rep}
2 S^\alpha_j=i \,\chi^\alpha_j \chi^0_j;\quad\alpha=x,y,z.
\end{alignat}
This representation induces a redundancy in the description, and in order to correctly reproduce the Hilbert space of a spin-$1/2$ particle, the constraint 
\begin{alignat}{1}\label{constraint}
\chi^x_i \chi_i^y \chi_i^z \chi^0_i=1
\end{alignat}
has to be implemented $\forall\,i$, wherefore
\begin{alignat}{1}\label{majorana rep full}
2 S^\alpha=i \chi^\alpha \chi^0=-i\frac{\epsilon_{\alpha\beta\gamma}}2\chi^\beta \chi^\gamma.
\end{alignat}
In this formulation, the model \eqref{eq:kitaev} can be kneaded into
\begin{alignat}{1}\label{eq:zero field}
H^{}_{\textsc{k}} = -\frac{1}{4}\sum_{\langle i,j\rangle}\left(i K_{\alpha_{ij}}\,\chi_i^{\alpha_{ij}}\chi_j^{\alpha_{ij}}\right)\left(i\, \chi^0_i \,\chi^0_j\right).
\end{alignat}
Physically, this can be thought of as a simple problem of Majorana fermions $\{\chi^0_i\}$ that hop with a bond-dependent amplitude $t_{ij}$\,$=$\,$i K_{\alpha_{ij}}\chi_i^{\alpha_{ij}}\chi_j^{\alpha_{ij}}$. A key observation by \citet{kitaev2006anyons} was that the $\mathbb{Z}_2$ flux around each hexagonal plaquette $p$,
\begin{alignat}{1}\label{z2 flux}
\Phi_p=\prod_{\langle i,j\rangle\,\in\, p}(i \chi_i^{\alpha_{ij}}\chi_j^{\alpha_{ij}})= \pm1,
\end{alignat}
is a conserved quantity at zero field. Therefore, Eq.~\eqref{eq:zero field} can be reinterpreted as describing Majorana fermions hopping in a background $\mathbb{Z}_2$ gauge flux. Conveniently, Lieb's theorem \cite{lieb1994} then asserts that the ground state is given by a  uniform zero-flux state where $\Phi_p = +1$ $\forall\,p$. 

Solving the Bogoliubov-de Gennes (BdG) Hamiltonian for $\{\chi^0_i\}$ in momentum space leads to a fermionic band structure, which encodes all the information about the (short-ranged) spin correlations \cite{baskaran2007exact}. The spectrum is fully gapped if $\lvert K_z\rvert$\,$>$\,$\lvert K_x\rvert $\,$+$\,$\lvert K_y\rvert$, which places us in the $A$ phase. Contrarily, if $\lvert K_z\rvert$\,$<$\,$\lvert K_x\rvert$\,$+$\,$\lvert K_y\rvert$, one finds a graphene-like band structure with a pair of Dirac points at zero energy, positioned at momenta $\pm\arccos \,[-K_z/(2K)]$ (taking $K_x$\,$=$\,$K_y$\,$\equiv$\,$K$). This corresponds to the phase $B$, which carries gapped vortices and gapless fermions. A low-energy description of this phase is thus given by Dirac fermions coupled to a dynamical $\mathbb{Z}_2$ gauge field.

Moving away from the solvable limit, we now add to $H_\textsc{k}$ a Zeeman coupling to the magnetic field
\begin{equation}
\label{eq:Bfield}
H^{}_\textsc{z} = - \sum_{j} \vec{h}\cdot \vec{S}_j = - \sum_{i} \left(h_x S^x_j + h_y S^y_j + h_z S^z_j \right),
\end{equation}
where $\vec{h}$\,$\equiv$\,$(h_x, h_y,h_z)$ is the applied field, and we have absorbed the Bohr magneton $\mu_B$ in its definition.
\citet{kitaev2006anyons} proved that a generic perturbation of this kind opens up a spectral gap in the originally gapless $B$ phase.
To see this, we can consider the effect of $H_\textsc{z}$ in perturbation theory within the zero-flux (or vortex-free) sector; for simplicity, let us assume isotropy, i.e., $K_x=K_y=K_z \equiv K$. 
In this low-energy sector, all perturbations vanish at first order in $h$, while the second-order terms simply renormalize the original coupling $K$ between each nearest neighbor (abbreviated hereafter as NN).
The lowest nonzero correction actually arises at third order in the field, leading to an effective Hamiltonian
\begin{equation}
\label{eq:three-spin}
H^{}_{3s} \simeq - \frac{h_x\, h_y\, h_z}{K^2} \sum_{j,k,l}S^x_j S^y_k S^z_l,
\end{equation}
where the summation runs over two possible configurations of three spins arranged as follows:
\begin{equation*}
\includegraphics[width=0.95\linewidth,trim={0 0.25cm 0 0.25cm}, clip]{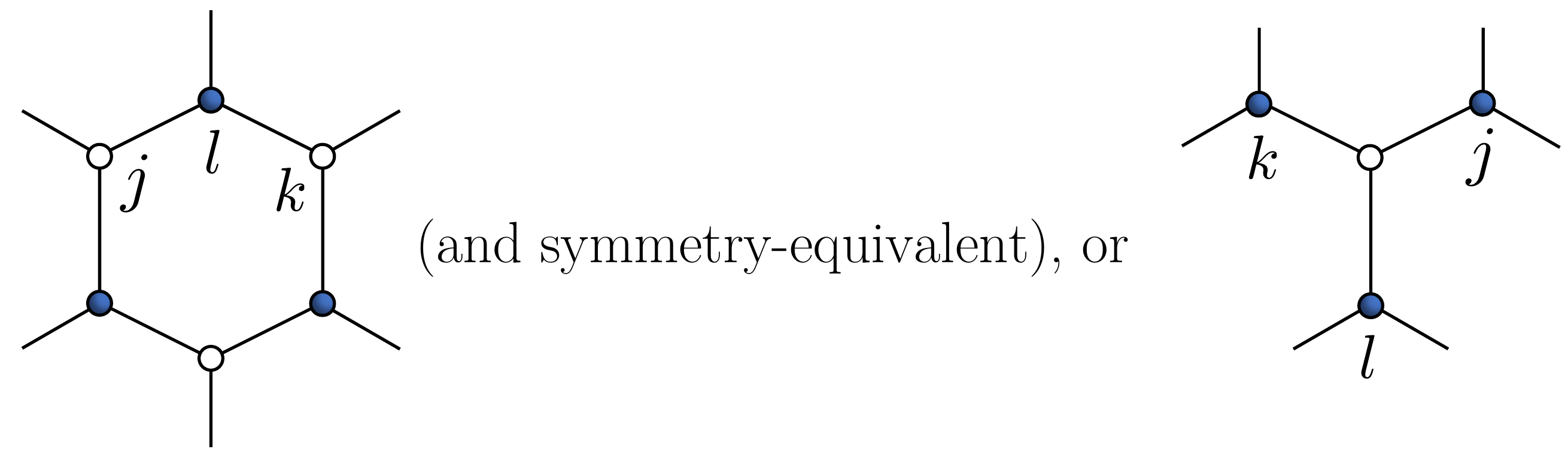}.
\end{equation*}
In terms of the Majorana fermions $\{\chi^0_i\}$, $H_{3s}$ generates second-NN hopping as well as four-fermion interactions, and introduces a gap in the spectrum.

Taken together, $H_\textsc{k}+H_\textsc{z}+H_{3s}$ now encompasses all the ingredients for a mean-field Hamiltonian of the pure Kitaev model \eqref{eq:kitaev} coupled to a magnetic field, which we will construct in Sec.~\ref{sec:mf}. Since $H_{3s}$ is derived above as only a perturbative approximation to $H_\textsc{z}$, including both these terms in a theory might naively seem redundant. However, at the mean-field level,  we allow for all possible symmetry-permitted terms, so it becomes necessary to separately incorporate the distinct first- and second-NN interactions stemming from Eqs.~\eqref{eq:Bfield} and \eqref{eq:three-spin}, respectively.

\subsection{Symmetries}
\label{sec:symm}

To proceed further, we have to establish the space group symmetries of the Kitaev model that must be taken into account by our eventual mean-field theory. In this regard, it is useful to visualize the honeycomb lattice in Fig.~\ref{fig:symmetry}(a) as being embedded within a 3D cubic lattice [Fig.~\ref{fig:symmetry}(b)], as is indeed the case in material realizations such as the layered iridates \cite{chaloupka2010kitaev, singh2010, singh2012, rau2016spin}. Given the strong spin-orbit coupling, all symmetry transformations must act simultaneously on the spin and spatial degrees of freedom, which live in \textit{three-dimensional} real space (rather than on the 2D honeycomb lattice alone). 

In the absence of a magnetic field, the space group is generated by the following elements. Firstly, the Hamiltonian enjoys the translational symmetries $T_{1,2}$ along the two primitive lattice vectors $\vec{n}_{1,2}$. One possible set of point-group generators is \cite{you2012doping, ZouandHe2020}:\\
1. Inversion---or twofold rotation---$C_2$; the representation of this symmetry is simply
\begin{equation}
C^{}_2:\,\, S_{\vec r}^x \rightarrow S_{C_2 \vec r}^x, \,\,
 S_{\vec r}^y \rightarrow S_{C_2 \vec r}^y, \,\,
 S_{\vec r}^z \rightarrow S_{C_2 \vec r}^z.
\end{equation}
2. Pseudo-mirror $R^*$, composed of the conventional mirror symmetry (namely, a reflection $R$ across the $x$\,$=$\,$y$ plane) and a spin rotation  $\mathrm{e}^{i \pi S^y}$\,$\mathrm{e}^{i (\pi/2) S^z}$, which acts as 
\begin{equation}
R^*: \,\,  S_{\vec r}^x \rightarrow -S_{R \vec r}^y, \,\,
 S_{\vec r}^y \rightarrow -S_{R \vec r}^x, \,\,
 S_{\vec r}^z \rightarrow -S_{R \vec r}^z.
\end{equation}
3. Improper rotation $S_6$, defined by a sixfold rotation about the $c$ axis, followed by a reflection across the $c$\,$=$\,$0$ lattice plane i.e., $S_6$\,$\equiv$\,$C_6 \cdot \mathrm{e}^{i (2\pi/3) (S^x+S^y+S^z)/\sqrt{3}}$ such that $(S_6)^6$\,$=$\,$1$. This symmetry holds only for the isotropic Kitaev model with a Zeeman field in the $[1,1,1]$ direction.\\
The components of these three operations acting on the 2D honeycomb lattice are sketched in Fig.~\ref{fig:symmetry}

In addition, the zero-field Kitaev model naturally possesses time-reversal symmetry. The antiunitary time-reversal operation ($\Theta$) has no effect on the lattice per se but acts on the spins as $i S^y\, \mathbb{K} $, where $\mathbb{K}$ denotes complex conjugation. Even though $\Theta^2=-1$ for a single spin, note that we have $\Theta^2=+1$ for the global time-reversal symmetry operation due to the bipartite nature of the honeycomb lattice.

The time-reversal symmetry will, of course, be broken by a finite magnetic field. Furthermore, a field along a generic direction also breaks the pseudo-mirror symmetry. Both these properties of the applied field are crucial since the presence of either time-reversal or pseudo-mirror symmetry prohibits a finite thermal Hall conductivity. We can illustrate this point by contrasting two specific field directions. Let $\kappa_{xy}$ denote the in-plane thermal Hall conductivity, with both the temperature gradient and the ensuing heat current in the honeycomb-lattice planes depicted in Fig.~\ref{fig:symmetry}(a). Now, for example, if we take $\vec{h} \,\|\, [\bar{1}1 0]$, parallel to the $b$ axis, then $\kappa_{xy}$ must necessarily vanish as a consequence of the $R^*$ symmetry. On the contrary, if $\vec{h}\, \|\, [11x]$, in the $ac$ plane, then the pseudo-mirror and time-reversal symmetries are individually broken but their combination is preserved; in this case, one can have a nonzero $\kappa_{xy}$. Hence, in experiments \cite{kasahara2018majorana}, the Zeeman field is aligned to be on the $ac$ plane. We will begin by considering a magnetic field along the $[1 1 1]$ direction; thereafter, we generalize the orientation to $[11x]$ and observe the change in the thermal Hall response brought about by such a rotation.

\subsection{Parton construction}
\label{sec:parton}

While Kitaev's original solution of the model \eqref{eq:kitaev} entailed a rewriting of the spin variables in terms of Majorana fermions, the correct low-energy degrees of freedom can also be singled out by a different  fermionization procedure using spinful complex fermions \cite{burnell20112}. 
Guided by this correspondence, we will use the latter formalism to study the Kitaev spin liquid and proximate phases upon perturbing away from the exactly solvable zero-field limit.

In the Abrikosov fermion representation \cite{abrikosov1965electron, affleck19882, marston1989large} motivated above, the spin operator at each site is decomposed as:
\begin{equation} \label{eq:parton}
\vec S^{}_i=\frac{1}{2}c_i^\dag\,\vec\sigma \,c_i^\pdagger;
\end{equation}
here, $c_i \equiv (c_{i,1}, c_{i,2})^\mathrm{T}$ is a two-component fermionic spinon operator, and $\vec\sigma$ denotes the three usual Pauli matrices. Importantly, the mapping from spin-$1/2$ to fermions in \equref{eq:parton} expands the Hilbert space and, in order to remain within the physical subspace, we must restrict ourselves to the fermionic states with single occupation per site. Hence, this decoupling is to be supplemented with the constraints
\begin{equation} \label{eq: gauge constraint}
c_i^\dag c^\pdagger_i=1,\,\, c_{i,1}^\dag c_{i,2}^\dag = 0,\,\, c_{i,1}^\pdagger c_{i,2}^\pdagger =0,
\end{equation}
and therefore, any faithful fermionic band structure of the spinons is always constrained to be at half-filling.

Related to this constraint, the parton construction outlined above exhibits an SU(2) gauge structure \cite{lee2006doping, hermele2007}. This can be made apparent by defining the matrix
\begin{equation}
\mc{C}_i=
\left(
\begin{array}{cc}
	c^\pdagger_{i,1} & -c_{i,2}^\dag\\
	c^\pdagger_{i,2} & c_{i,1}^\dag
\end{array}
\right)\label{SpinonMatrix}
\end{equation}
containing the spinon operators on site $i$. The physical spin operators can now be written in terms of $\mc{C}_i$ as
\begin{equation} \label{eq: parton-usual-2}
\vec S^\pdagger_i=\frac{1}{4}\mathrm{Tr}\left(\mc{C}_i^\dag\,\vec\sigma \,\mc{C}^\pdagger_i\right).
\end{equation}
As (\ref{eq: parton-usual-2}) is invariant under a local SU$(2)$ transformation
\begin{equation} \label{eq: SU(2)-redundancy}
\mc{C}^\pdagger_i\rightarrow \mc{C}^\pdagger_i W^\pdagger_i,
\end{equation}
where $W_i$ is an SU$(2)$ matrix, this parton construction has an SU$(2)$ gauge redundancy.
This leads to a description of the underlying spin model as a theory of fermions coupled to an SU(2) gauge field \cite{affleck19882, coleman1988kondo, andrei1989cooper}. However, the actual residual gauge group can be smaller than the full SU(2) depending on the particular phase of interest. For instance, in the SFS state, the SU$(2)$ symmetry is broken down to the U$(1)$ subgroup---this is an emergent dynamical U$(1)$ gauge field (as opposed to the conventional U$(1)$ electromagnetic field under which the spinons are charge-neutral); the associated gauge transformation that leaves the spins invariant reads as $c_i\rightarrow c_i \mathrm{e}^{i\theta_i}$. Moreover, if the spinons are in a superconducting phase (such as in the CSL), then this gauge symmetry is broken down to $\mathbb{Z}_2$ ($c_i\rightarrow \pm c_i$) by the pairing terms.


Owing to the gauge redundancy arising from the Abrikosov fermion representation, a gauge transformation $g\in\mathbb{G}$, with $\mathbb{G}$ being the residual gauge group, leaves the Hamiltonian invariant. Any operation---including, in particular, the symmetry transformations listed in Sec.~\ref{sec:symm}---can act within this gauge space in addition to the spin degrees of freedom. Hence, all symmetries act \textit{projectively} and are defined by the corresponding left ($W$) and right ($G$) multiplications of the spinon matrix $\mc{C}$ in \equref{SpinonMatrix}: this information, known as the projective symmetry group (PSG) \cite{wen2002quantum, essin2013classifying}, characterizes the fractionalized phases. The PSG for the Kitaev model was worked out by Ref.~\cite{you2012doping}. 
In a generic gauge, $c$ transforms to a linear combination of $c$ and $c^\dagger$. 
Such a description is inconvenient for U(1) SFS spin liquids, as it would imply that pairing terms ($c_i c_j$ + h.c.) could be generated from purely hopping terms ($c^\dagger_i c^\pdagger_j$ + h.c.) due to symmetry transformations alone. 
This drawback can be circumvented, however, by choosing a suitable gauge \cite{ZouandHe2020}. We define such a gauge in Appendix~\ref{app:kitaev_f_H} and denote the corresponding spinon operators by $f^\pdagger_{i\eta},\,f^\dagger_{i\eta}$, $\eta=1,2$, in the following.
In that gauge, the symmetries act as
\begin{subequations}
\label{eq:symm}
\begin{alignat}{1}
T_{1, 2}&:	 f_i\rightarrow  f_{i+\vec n^{}_{1, 2}}\\
S_6&:  f_{i}\rightarrow \mathrm{e}^{i\frac{5\pi}{6}}\,U_{S_6}^\dag \, f_{C^{}_6 i}\\
\Theta R^*&:  f_{i}\rightarrow  \mathrm{e}^{-i\frac{\pi}{4}}\,U_{\Theta R^*}^\dag\,  f_{R\, i},
\end{alignat}
\end{subequations}
where
\begin{subequations}
\begin{alignat}{1}
U^\pdagger_{S_6} &\equiv\frac{1+i(\sigma_1+\sigma_2+\sigma_3)}{2}\\ 
U^\pdagger_{\Theta R^*} &\equiv \mathrm{e}^{-i\sigma_3\frac{\pi}{4}}
\end{alignat}
\end{subequations}
Now we can see explicitly that the gauge charge of the spinons is preserved by the symmetry implementation. This will be very convenient in the following mean-field treatment.

\subsection{Mean-field theory}
\label{sec:mf}

The Kitaev honeycomb lattice model was studied using the SU(2) fermion formalism by Ref.~\cite{burnell20112}, which showed that the description of Ref.~\cite{kitaev2006anyons} can be exactly reproduced in this language. To be precise, the physical correlation functions of the true ground state of Eq.~\eqref{eq:kitaev} are captured by a stable mean-field theory which can be constructed as follows.

Since all the Kitaev interactions in $H_\textsc{k}$ involve two spins, inserting the representation \eqref{eq:parton} generates terms that are a product of four fermions; the resultant fermionic Hamiltonian is rather complicated due to the lack of spin rotation invariance in the model. One way to proceed is to use a Hubbard-Stratonovich transformation \cite{hubbard1959} to decouple the four-fermion interactions, which can be recast into interactions between a pair of fermion operators on the sites $i$ and $j$ and a
bosonic field (which lives on the link between them). At the mean-field level, these four auxiliary fields assume nonzero expectation values. Imposing the self-consistency of the expectation values, which can be expressed in terms of $K_{x,y,z}$ leads to the coefficients of the quadratic terms of the Hamiltonian at the saddle point of interest.
 
However, significant physical insight can be gleaned from a phenomenological analysis of such a mean-field description without necessarily self-consistently solving the theory. To this end, we rewrite the Kitaev model in terms of the spinon operators $f$ as the sum:
\begin{equation} \label{eq: Kitaev parton}
H_\textsc{k}^{\textsc{mf}}=H^{}_{\rm hopping}+H^{}_{\rm pairing}.
\end{equation}
The detailed form of these terms are documented in Appendix \ref{app:kitaev_f_H}. $H_{\rm hopping}$ consists solely of hopping operators of the spinons, \ie each term therein preserves the U$(1)$ symmetry. Conversely, $H_{\rm pairing}$ contains purely pairing terms of the spinons that break the U$(1)$ gauge symmetry down to $\mathbb{Z}_2$. In total, $H_\textsc{k}^{\textsc{mf}}$ is described by  two types of first-NN interactions, of strengths $J_1$ and $J_1'$, and second-NN interactions with a coupling $J_2$. Eq.~\eqref{eq:three-spin} informs us that such a second-NN term originates from the effect of a magnetic field in third-order perturbation theory, so, in principle, $J_2$ should be varied as $\propto h_x h_y h_z$. This perturbative expansion, of course, only holds for small $h$; in the regime of large magnetic fields, we can think of a constant $J_2$ as being spontaneously induced by the field.

\begin{figure}[tb]
\includegraphics[width=\linewidth]{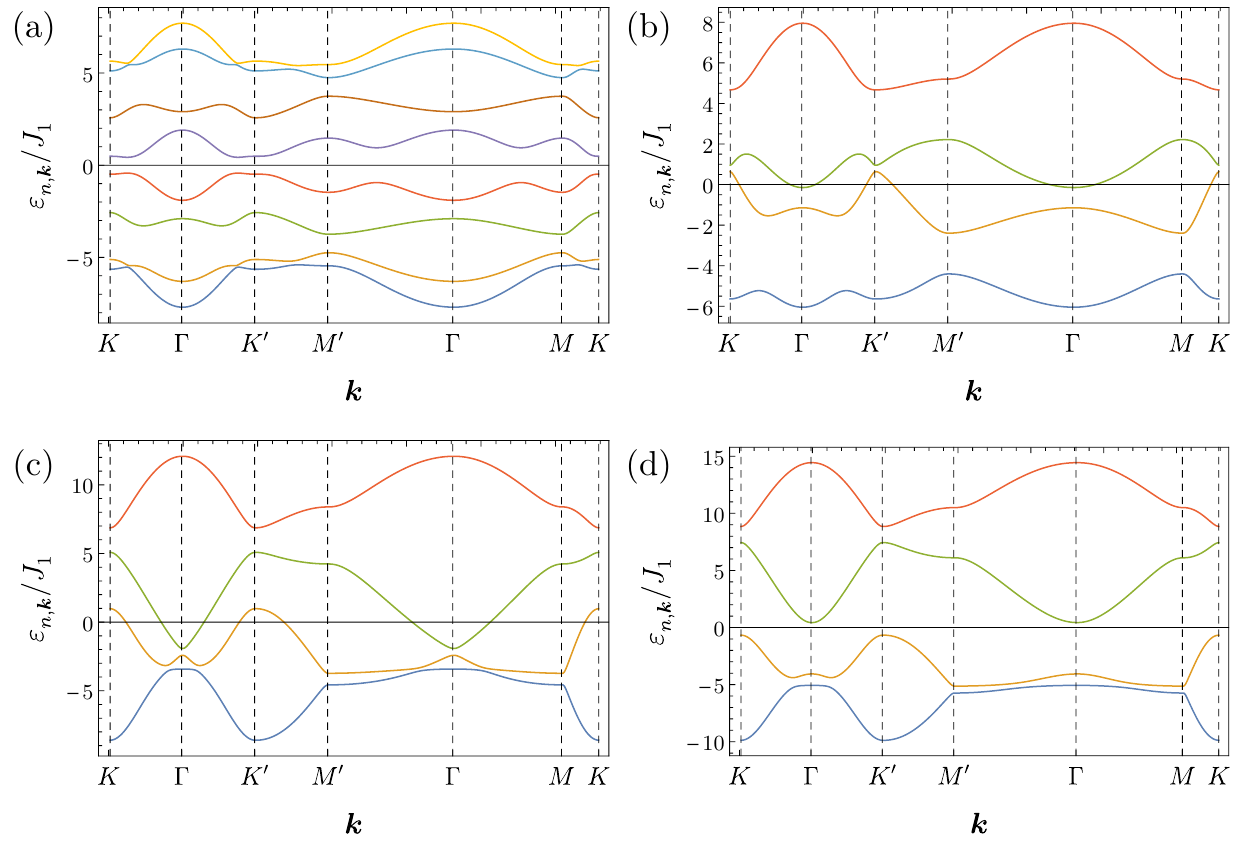}
\caption{Dispersions of the mean-field spinon Hamiltonian, Eq.~\eqref{eq: Kitaev parton}, for the Kitaev model with $J_1=1$ and $J_1'=3.5$; plotted here is $\varepsilon_{n,\vec{k}}$\,$\equiv$\,$E_{n,\vec{k}}$\,$-$\,$\mu$. At low fields (a,b), we take the coefficient of the second-NN hopping to be $J_2$\,$=$\,$-2.5 h^3$, in accordance with Eq.~\eqref{eq:three-spin}. For $h$\,$=$\,$0.7/\sqrt{3}$ (a), the system is in the ITO phase; diagonalizing $H_{\rm ITO}$ [Eq.~\eqref{eq:H_ITO}] yields eight fully gapped bands. At $h$\,$=$\,$0.8/\sqrt{3}$ (b), the band structure clearly displays both electron-like and hole-like Fermi surfaces as expected in the SFS phase. When $h>1$ (c,d), $H_{3s}$ is no longer perturbative, so we employ an ansatz in which $J_2=-0.75$ and constant; the corresponding field strengths are (c) $h$\,$=$\,$4/\sqrt{3}$, and (d) $h$\,$=$\,$6/\sqrt{3}$. The Fermi surfaces shrink as the field is increased, eventually leading to the gapped polarized phase (d).}
\label{fig:bands_kitaev}
\end{figure}

The mean field Hamiltonians that we construct for the different phases are thus
\begin{subequations}\label{HamiltonianBothPhases}
\begin{alignat}{1} 
\label{eq:H_ITO}
H^{}_{\rm ITO}&=H_{\rm hopping}+\xi (\vec{h})\, H_{\rm pairing}+H^{}_{\vec{h}}\\
H^{}_{\rm SFS}&=H^{}_{\rm hopping}+H^{}_{\vec{h}}
\label{eq:H_SFS}
\end{alignat}\end{subequations}
where
\begin{equation} \label{eq: mean field Zeeman}
H^{}_{\vec{h}}=-\sum_if_i^\dag\left( h^{}_x \sigma^{}_1+ h^{}_y \sigma^{}_2+ h^{}_z\sigma^{}_3\right)f^{}_i
\end{equation}
represents the Zeeman coupling to a magnetic field in the $[h_x h_y h_z]$ direction. Nevertheless, $H_{\vec{h}}$ should not be literally taken as the full effect of a Zeeman field, since the latter can also renormalize the parameters in the other terms of the Hamiltonian. 
The strength of the pairing $\xi (\vec{h})$ is modulated as $\xi(\vec{h})$\,$=$\,$(1-|\vec{h}|/h_{c_1}\hspace{-0.1em}(\hat{\vec{h}}))^{1/2}$\,$\in$\,$[0,1]$ such that it vanishes at the critical field $h_{c_1}\hspace{-0.1em}(\hat{\vec{h}})$, and $\xi(\vec{h})$\,$=$\,$0$ for $|\vec{h}|$\,$>$\,$h_{c_1}\hspace{-0.1em}(\hat{\vec{h}})$. Since the Higgs mechanism is responsible for the ITO to SFS phase transition, the choice of a 1/2 power law is in analogy to the pairing amplitude of a superconductor. In general, the critical field $h_{c_1}$ depends on the direction of the magnetic field, $\hat{\vec{h}}$\,$=$\,$\vec{h}/|\vec{h}|$. For concreteness, however, we will neglect this anisotropy and set $h_{c_1}\hspace{-0.1em}(\hat{\vec{h}})$\,$=$\,$0.8$ throughout this work. We reiterate that choosing a gauge that does not mix $f$ and $f^\dagger$ in the symmetry transformations (\ref{eq:symm}) is essential here since, otherwise, dropping the pairing term would break the symmetries of the system and, as such, not constitute a proper description of the ITO to SFS transition. Within our description, $H_{\rm hopping}$, $H_{\rm pairing}$, and  $H_{\vec{h}}$ separately preserve all symmetries in Eq.~\eqref{eq:symm}.

To match the conventions of the literature with the magnetic field along the [111] direction, we use $h$ not to denote the magnitude of $\vec{h}$, but instead to parameterize it as $\vec{h} = (h,h,h)^T$ for this specific orientation of the magnetic field. 


Some representative dispersions of this mean-field Hamiltonian are shown in Fig. \ref{fig:bands_kitaev}. As the magnetic field is increased, the Fermi pockets emerge, change in size and shape, and eventually disappear. In this process there can be Dirac crossings
of the four bands, but since such crossings always occur away from the chemical potential, they do not induce a phase transition \cite{ZouandHe2020}.

\section{Thermal Hall response in the field-induced phases}

Prior to delving into the thermal Hall response of $H_\textsc{k}^{\textsc{mf}}$, let us briefly recast the phase diagram of the finite-field Kitaev model in the language of fermionic spinons.
Firstly, the gapless $B$ phase, within this description, is a $p$-wave superconducting state of the spinons with zero-energy excitations at nodal points \cite{burnell20112}; these excitations, in turn, constitute a single Dirac fermion. The Majorana fermions of the solution in Sec.~\ref{sec:majorana} appear as the BdG quasiparticles of the superconducting state.   In the presence of a field, the order parameter acquires an $i p$ component (resulting in a weak pairing \cite{read2000paired} $p_x+i p_y$ chiral topological superconductor) and the Dirac fermion develops a mass. This leads to a gapped ITO phase---with a non-Abelian chiral QSL ground state---which remains stable for small magnetic fields $\vec{h} \,\|\,[111]$ and weak anisotropy, i.e., $K_z$\,$\simeq$\,$K_x$\,$=$\,$K_y$\,$=$\,$K$. Recognizing the correspondence to a $p_x +i p_y$ superconductor, it immediately follows that this state must break both time-reversal and mirror-reflection symmetries as asserted previously.

Bordering the ITO phase is the gapless U$(1)$ spin liquid, which can be interpreted as a spinon metal. It is characterized by both electron and hole Fermi surfaces of neutral spinons, coupled to a dynamical U$(1)$ gauge field. This phase persists up to intermediate magnetic fields and weak anisotropy. Increasing the $[111]$ field further shrinks these pockets, bringing us to the gapped polarized phase, which is just a band insulator of spinons. This is  a partially polarized magnetic phase \cite{patel2019magnetic} that is adiabatically connected to the trivial fully-polarized product state at high fields. As a function of the field, the magnetization monotonically increases toward its saturation value attained when all the spins are aligned along the $[111]$ direction. 

\begin{figure}[tb]
\includegraphics[width=\linewidth,trim={0.4cm 0.6cm 0.3cm 0.5cm},clip]{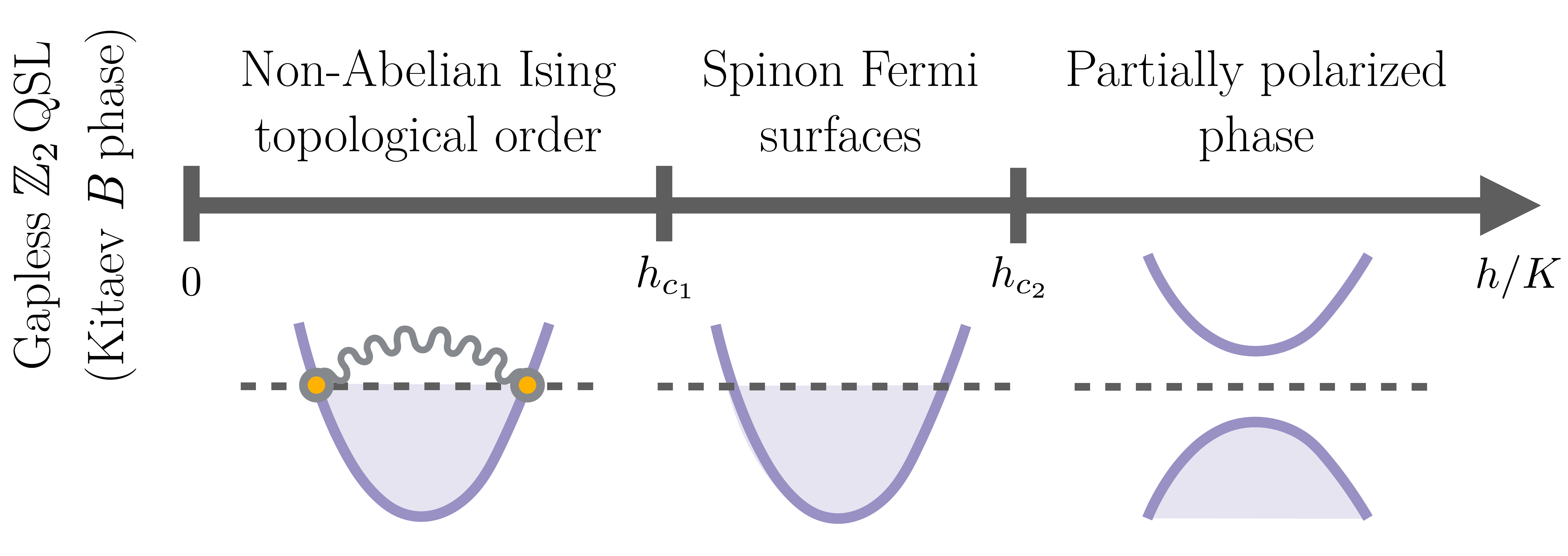}
\caption{Phases of the isotropic Kitaev model under a magnetic field in the $[111]$ direction, i.e., $h_x$\,$=$\,$h_y$\,$=$\,$h_z$\,$=$\,$h$. Starting with the $B$ phase at $h$\,$=$\,$0$, the system transitions from a non-Abelian chiral QSL with Ising topological order, to a gapless U(1) spin liquid, and finally, into a trivial polarized state as the field is varied. The three field-induced phases can also be reinterpreted in terms of the fermiology of the spinons as a $p_x+i \,p_y$ weak-pairing superconductor, a metal, and a band insulator, respectively, as depicted.}
\label{fig:phases}
\end{figure}

Finally, in the limit of strong anisotropy, one can also realize the $A$ phase of the Kitaev model. This gapped $\mathbb{Z}_2$ spin liquid corresponds to a trivial strong pairing $p$-wave superconductor of the spinons \cite{read1991large, Wen1991}. The state is fully gapped because the nodes in the order parameter do not intersect the Fermi surface. When $h\ll K$, the phase boundary between the non-Abelian ITO and this Abelian toric phase follows from perturbation theory \cite{jiang2018field} as
$K_z/K\approx2-38\,({h}/{K})^2+\mc{O}( h/ K)^4$.
Similarly, analyzing the properties of the toric code under a transverse field \cite{vidal2009low,vidal2009self,dusuel2011robustness}, the phase boundary with the polarized phase can be analytically determined to be $K_z/K\sim(h/K)^{-1} $. The line $K_z/K=1$, which we focus on, does not cross these boundaries in the $(h/K, K_z/K)$-plane, so we will never actually encounter the Abelian phase in our calculations. A schematic phase diagram summarizing the phases that we probe below is presented in Fig.~\ref{fig:phases}.

Our computation of the thermal Hall response will be carried out at the parton mean-field level, and we will not consider the consequences of gauge fluctuations. In the ITO, the gauge fluctuations are fully gapped and only lead to exponentially small corrections to $\kappa_{xy}$ at low temperatures. On the other hand, the U(1) gauge fluctuations in the gapless Fermi surface phase can lead to significant corrections: the structure of these corrections has been discussed elsewhere \cite{guo2020gauge}.

\subsection{Formalism}

The thermal Hall conductivity can be computed from the microscopic Hamiltonian using a linear response framework. It is, however, well recognized that calculations of $\kappa_{xy}$ based on a direct application of the Kubo formula are plagued by unphysical divergences at zero temperature \cite{katsura2010theory, matsumoto2011theoretical}. This is known to be a consequence of the broken time-reversal symmetry in the system.
Under such circumstances, a temperature gradient drives not only the transport current, but also an experimentally unobservable circulating current \cite{smrcka1977transport, cooper1997thermoelectric}. While the microscopic current density calculated by the standard linear response theory encapsulates both contributions, the circulating component has to be subtracted out for a well-defined response since it does not facilitate heat transport. As pointed out by  \citet{qin2011energy}, this can be achieved by carefully accounting for the electromagnetic and gravitomagnetic energy magnetizations \cite{luttinger1964theory, ryu2012electromagnetic}, which naturally arise as corrections to the thermal transport coefficients.

\begin{figure}[tb]
\includegraphics[width=\linewidth]{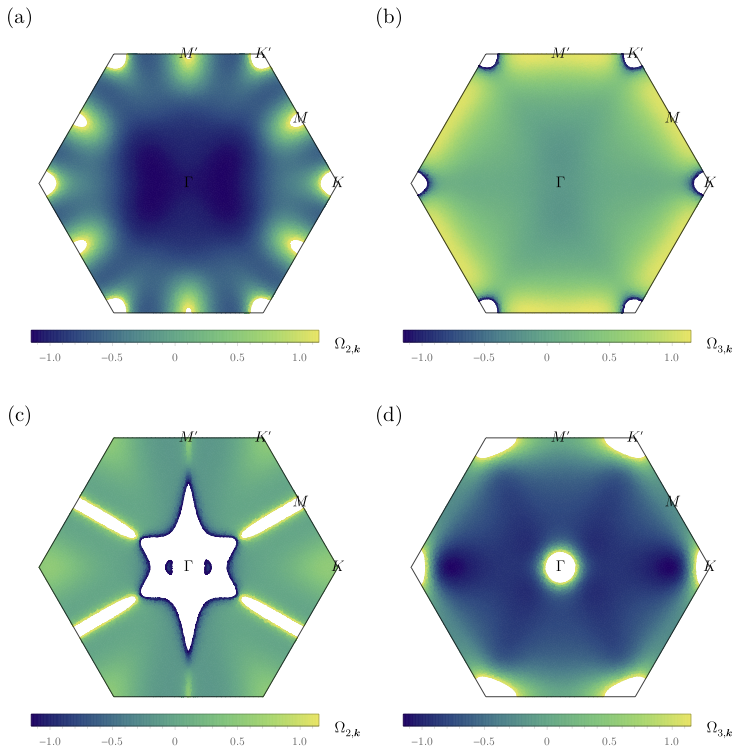}
\caption{Berry curvatures $\Omega_{n, \vec{k}}$ for the second, $n=2$, and third, $n=3$, bands (which cross the Fermi surfaces) in Fig.~\ref{fig:bands_kitaev}(b) and (c). The profiles here correspond to $\vec{h}=(h,h,h)^T$ with (a) $h$\,$=$\,$0.8/\sqrt{3}$, $n$\,$=$\,$2$, (b) $h$\,$=$\,$0.8/\sqrt{3}$, $n$\,$=$\,$3$, (c) $h$\,$=$\,$4/\sqrt{3}$, $n$\,$=$\,$2$, and (d) $h$\,$=$\,$4/\sqrt{3}$, $n$\,$=$\,$3$. The integrals of the curvatures over the Brillouin zone [Eq.~\eqref{eq:chern}] define the Chern numbers, which are (a) $2$, (b) $0$, (c) $-1$, and (d) $1$.}
\label{fig:berry}
\end{figure}

We now use the formalism of Ref.~\cite{qin2011energy} to first compute $\kappa_{xy}$ in the SFS phase, for which the Hamiltonian [Eq.~\eqref{eq:H_SFS}] does not involve any pairing terms. Transforming to momentum space, we have
\begin{equation}
f^\pdagger_{A(B),\eta} (\vec{k}) = \frac{1}{\sqrt{N}} \sum_{\vec{k}} \mathrm{e}^{-i \vec{k}\cdot\vec{r}}f^\pdagger_{A(B),\eta} (\vec{r}),
\end{equation}
where $A,B$ stand for the two sublattices, $N$ is the number of unit cells, and $\eta=1,2$. The mean-field Hamiltonian can now be expressed as
\begin{alignat}{1}
\label{eq:Hk1}
H^{\textsc{mf}}_{\textsc{k}} &= \sum_{\vec{k}}\psi^\dagger_{\vec{k}} \,H(\vec{k}) \,\psi^\pdagger_{\vec{k}},\\
\nonumber\psi^\pdagger_{\vec{k}} &= \left(f^\pdagger_{A,1} (\vec{k}), f^\pdagger_{A,2} (\vec{k}),f^\pdagger_{B,1} (\vec{k}),f^\pdagger_{B,2} (\vec{k})\right)^\mathrm{T}.
\end{alignat}
Diagonalizing $H(\vec{k})$ produces the dispersion in Fig.~\ref{fig:bands_kitaev} (b-d); there are four bands, labeled by $n$, with the corresponding eigenergies $E_{n, \mathbf{k}}$. The thermal transport coefficient is directly related to the Berry curvature in momentum space \cite{katsura2010theory}, which is given by
\begin{equation}
\Omega^{}_{n, \vec{k}} = - 2\, \mathrm{Im} \left \langle \frac{\partial \,u^{}_{n, \vec{k}}}{\partial k^{}_x } \bigg\vert \frac{\partial\, u^{}_{n, \vec{k}}}{\partial k^{}_y } \right \rangle,
\end{equation}
$u_{n,\vec{k}}$ being the periodic part of the Bloch wavefunction with band index $n= 1,\ldots,4$. For reference, Fig.~\ref{fig:berry} displays $\Omega_{n, \vec{k}}$ for the $n$\,$=$\,$2$ and $n$\,$=$\,$3$ bands (colored yellow and green, respectively) of Fig.~\ref{fig:bands_kitaev}(b) and (c), corresponding to the SFS phase at two different fields; we will see shortly how the variations in $\kappa_{xy}$ can be connected to the momentum space distribution of the Berry curvatures.
Defining
\begin{equation}
\label{eq:curvature}
\sigma^{}_{xy} (\epsilon) = -\int_{E_{n, \mathbf{k}}< \epsilon} \frac{\mathrm{d}^2 \vec{k}}{(2 \pi)^2} \,\, \Omega_{n, \vec{k}},
\end{equation} 
which is simply $\hbar/e^2$ times the zero-temperature
anomalous Hall coefficient for a system with chemical potential $\epsilon$ \cite{xiao2006berry, jungwirth2002anomalous}, the thermal Hall conductivity is given by  \cite{qin2011energy}
\begin{equation}
    \kappa^{}_{xy} = -\frac{k_B^2}{\hbar\, T} \int \mathrm{d}\epsilon\, (\epsilon-\mu)^2\, \sigma^{}_{xy}\, (\epsilon)\, \mathfrak{f}'(\epsilon-\mu)
    \label{eq:k_xy}
\end{equation}
where $\mu$ is the chemical potential and $\mathfrak{f}\,(\epsilon)$ is the Fermi distribution function. Enforcing the parton constraint in Eq.~\eqref{eq: gauge constraint} fixes $\mu$ so that the system is always maintained at half-filling.

For an isolated band separated from all others by an energy gap, the Chern number, which is the integral of the Berry curvature over the Brillouin zone,
\begin{equation}
\label{eq:chern}
C^{}_n = \frac{1}{2\pi} \int \mathrm{d}^2 k \,\, \Omega^{}_{n, {\vec{k}}} \in \mathbb{Z},
\end{equation}
is well-defined and integer-valued. Using the Sommerfeld expansion, it is easy to see that as $T \rightarrow 0$, 
\begin{equation}
    \frac{\kappa^{}_{xy}}{T} = - \frac{\pi\, k_B^2}{6\, \hbar} \sum_{n \, \in \, {\rm filled~bands}} \hspace*{-0.3cm}C^{}_n.
    \label{eq:k_xy0T}
\end{equation}
Consequently, $\kappa_{xy}/T$ is quantized in units of $\pi/6$ as $T$\,$\rightarrow$\,$0$. On the other hand, if either the occupied bands are all topologically trivial or the net sum of their Chern numbers is zero, then $\kappa_{xy}$ eventually vanishes at $T$\,$=$\,$0$. Clearly, this analysis is not applicable to the SFS state, which is gapless, but it will prove to be relevant to the polarized phase, as well as to Sec.~\ref{sec:th_h} below.

\begin{figure*}[htb]
\includegraphics[width=\linewidth]{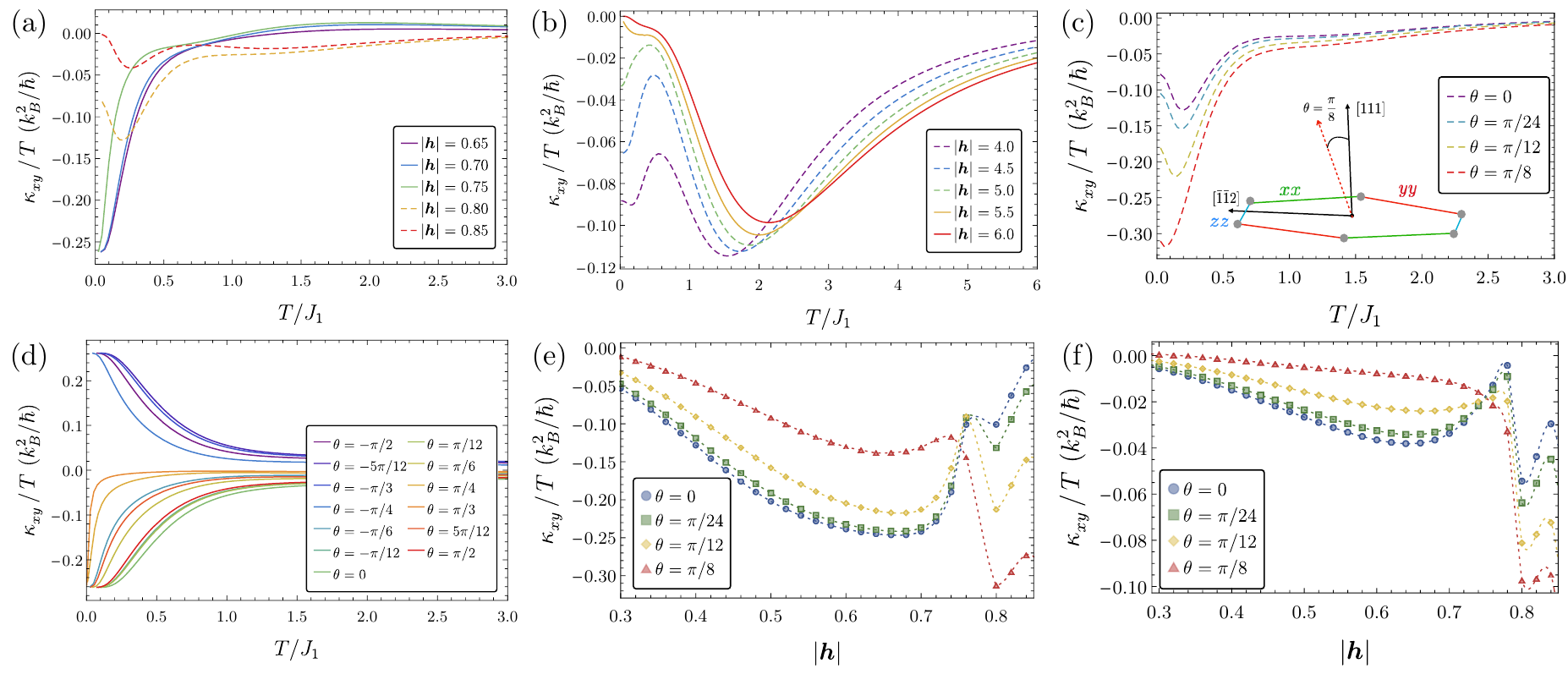}
\caption{\label{fig:th_kitaev}Thermal Hall conductivity of the Kitaev model (in mean-field theory) as a function of field strengths and orientations. Gapped (gapless) phases are portrayed as solid (dashed) lines. (a) For small magnetic fields in the $[111]$ direction, the zero-temperature value of $\kappa_{xy}/T$ is half-quantized at $-\pi/12$, which is indicative of the Majorana edge state. Increasing $|\vec{h}|$ leads to the onset of the SFS phase, whereupon we obtain an unquantized coefficient. (b) The SFS phase transitions to the trivial partially polarized state at high fields; the latter is characterized by a vanishing thermal Hall response as $T$\,$\rightarrow$\,$0$. (c) Temperature dependence of $\kappa_{xy}/T$ for a magnetic field in the $[11x]$ direction with a fixed magnitude $\lvert \vec{h} \rvert$\,$=$\,$0.8$ but variable polar angle $\theta$ with respect to the $c$-axis in the $ac$ plane. (d) Scanning the angle from $\theta$\,$=$\,$-\pi/2$ to $\theta$\,$=$\,$\pi/2$ in increments of $\pi/12$ for a magnetic field $|\vec{h}|$ = 0.6 shows clear anisotropy, with a change of sign between $\theta$\,$=$\,$-\pi/4$ and $\theta$\,$=$\,$-\pi/6$.  (e--f) Field dependence of $\kappa_{xy}/T$ for a magnetic field in the $[11x]$ direction with fixed temperatures $T/J_1$\,$=$\,$0.1$ (e) and $T/J_1$\,$=$\,$0.5$ (f).}
\label{fig:k_xy0T}
\end{figure*}

The story has to be modified when the Hamiltonian involves spinon pairing terms, such as for the ITO phase [Eq.~\eqref{eq:H_ITO}], since the spinor structure in Eq.~\eqref{eq:Hk1} can no longer be used. Instead, we can manipulate the Hamiltonian into the BdG form,
\begin{alignat}{1}
\label{eq:HBdG}
H^{\textsc{mf}}_{\textsc{k}} &= \frac{1}{2}\sum_{\vec{k}}\Psi^\dagger_{\vec{k}} \,H^\pdagger_{\mathrm{BdG}}(\vec{k}) \,\Psi^\pdagger_{\vec{k}},\\
\nonumber\Psi^\pdagger_{\vec{k}} &= \left(f^\pdagger_{A} (\vec{k}), f^\pdagger_{B} (\vec{k}),f^\dagger_{A} (-\vec{k}),f^\dagger_{B} (-\vec{k})\right)^\mathrm{T}
\end{alignat}
where $\Psi$ is an eight-component Nambu spinor; note that we have suppressed the index $\eta = 1,2$ on each of $f_{A,B}$ for brevity of notation. The bands obtained upon diagonalization of $H_{\mathrm{BdG}}$ are plotted in Fig.~\ref{fig:bands_kitaev}(a); we, once again, denote the associated eigenenergies and wavefunctions by $E_{n,\vec{k}}$ and $u_{n,\vec{k}}$, respectively, but with the distinction that $n=1,\ldots,8$. This eight-band description necessitates a theory of the thermal Hall effect for superconductors \cite{vafek2001quasiparticle, wang2011topological, nomura2012cross,stone2012gravitational}.
The most general formalism in this regard was developed by Ref.~\cite{sumiyoshi2013quantum}, starting from the assumptions that the BdG Hamiltonian is Hermitian and preserves particle-hole symmetry, both of which are satisfied by Eq.~\eqref{eq:HBdG}.
The end result is remarkably simple:
\begin{alignat}{1}
    \label{eq:k_xy2}
    \kappa^{}_{xy} &= -\frac{1}{2}\frac{k_B^2}{\hbar\, T} \int \mathrm{d}\epsilon\, (\epsilon-\mu)^2\, \sigma^{}_{xy}\, (\epsilon)\, \mathfrak{f}'(\epsilon-\mu),\\
        \lim_{T\rightarrow 0} \frac{\kappa^{}_{xy}}{T}& = - \frac{\pi\, k_B^2}{12\, \hbar} \sum_{n\,\vert\, E_{n,\vec{k}} \le 0} C^{}_n,
\label{eq:k_xy0T2}
\end{alignat}
where $\sigma$ and $C_n$ are defined exactly as in Eqs.~\eqref{eq:curvature} and \eqref{eq:chern}, respectively, but for the BdG spectrum. The crucial difference compared to Eq.~\eqref{eq:k_xy0T} is the additional factor of $1/2$, which implies that $\kappa_{xy}/T$ is now half-quantized at zero temperature. Nonetheless, if all pairing terms were to be dropped, Eqs.~\eqref{eq:k_xy} and \eqref{eq:k_xy2} would yield identical answers for $\kappa_{xy}$.

Notably, the derivation of the quantum thermal Hall conductivity in Ref.~\cite{sumiyoshi2013quantum} relies solely on the bulk microscopic Hamiltonian without any reference to the edge whatsoever. However, their final formula [Eq.~\eqref{eq:k_xy2}] is in complete agreement with the result in Ref.~\cite{read2000paired}, which studied the purely edge theory in a spinless chiral $p$-wave superconductor to show that the thermal Hall coefficient in the low-temperature limit is precisely $c\,(\pi T/6) (k_B^2/\hbar)$, where $c=1/2$ is the central charge of the Ising conformal field theory describing the Majorana edge state.

\subsection{Results}
\label{sec:results_kitaev}

The thermal Hall conductivity for the parton mean-field theory [Eq.~\eqref{eq: Kitaev parton}]  of the Kitaev model is shown in Fig.~\ref{fig:th_kitaev}. Let us first concentrate on the low-field regime with $\vec{h}$\,$\|$\,$[111]$: under these conditions, as discussed earlier, the second-NN interactions can be regarded as arising from the perturbative three-spin term of Eq.~\eqref{eq:three-spin}, so we set $J_2$\,$=$\,$-2.5 h^3$. Figure~\ref{fig:th_kitaev}(a) illustrates that for small $h$, $\kappa_{xy}/T$ is quantized at precisely $-\pi/12$ as $T\rightarrow 0$ over a substantial field range. This plateau, which is at half of the two-dimensional thermal Hall conductance in integer quantum Hall systems, agrees perfectly with the existence of the Majorana edge mode in the ITO phase. The negative sign simply follows from Eq.~\eqref{eq:k_xy0T2} as the Chern numbers of the lowest four bands in Fig.~\ref{fig:bands_kitaev}(a) are $\{1,-1,0,1 \}$ and thus, sum to $+1$. In the opposite limit of high temperatures ($T$\,$\gg$\,$J_1$), the bands are all equally populated, as determined by the Fermi distribution function; since the net Chern number of all the bands is necessarily zero, the thermal Hall conductivity also vanishes as $T$\,$\rightarrow$\,$\infty$. The quantization persists up to $|\vec{h}| \le h_{c_1} = 0.8$. 
Increasing the field beyond this critical value results in the formation of Fermi surfaces, as seen in Fig.~\ref{fig:bands_kitaev}(b), and the response ceases to be pinned at $-\pi/12$. This transition from the ITO to the SFS phase is described by QCD$_3$-Chern-Simons theory, which has emergent gapless Dirac fermions (with $N_f$\,$=$\,$1$ flavors) coupled to a U($2$) Chern-Simons gauge field \cite{ZouandHe2020,hsin2016level,seiberg2016gapped}. Once the Fermi surfaces begin to grow, there is an additional component to the zero-temperature value of $\kappa_{xy}/T$, which we can quantify as  $\Delta$\,$\equiv$\,$\lim_{T\rightarrow 0}\, (\kappa_{xy}/T)$\,$-$\,$(-\pi/12)$. The sign of this deviation $\Delta$ is positive in Fig.~\ref{fig:th_kitaev}(a) and can be understood in terms of the Berry curvatures as follows. As $h$ is increased from $h_{c_1}$, the electron-like Fermi surfaces near the $\Gamma$ point of the Brillouin zone, as well as the hole-like pockets near the $K$ and $K'$ points, start to expand. In the process, $\kappa_{xy}$ effectively gains (loses) some contribution from part of the third (second) band.
However, from Fig.~\ref{fig:berry}(b), we can see that the Berry curvature of the $n$\,$=$\,$3$ band centered around the $\Gamma$ point is negative, so, by Eq.~\eqref{eq:k_xy}, the portion of the third band below the Fermi surface contributes to a positive $\Delta$. Analogously, the curvature of the $n$\,$=$\,$2$ band in the vicinity of $K$ and $K'$ is positive, and therefore, given the hole-like nature of the Fermi surfaces concerned, this too leads to a net positive $\Delta$.

Proceeding to even larger $h$, one would expect to move beyond the scope of Eq.~\eqref{eq:three-spin}; therefore, we now take the parameter $J_2$ to be a constant (instead of the earlier cubic $h$ dependence); physically, this amounts to asserting that a finite $J_2$ emerges spontaneously as the effect of an external magnetic field. Specifically, we set the ratio $J_2/J_1$\,$=$\,$-0.75$, as suggested by the density-matrix renormalization group (DMRG) results of \citet{ZouandHe2020}. Upon increasing the field, the Fermi surfaces gradually shrink [see Fig.~\ref{fig:bands_kitaev}(c)], and the system is driven into the field-induced polarized phase. Figure~\ref{fig:th_kitaev}(b) highlights the thermal Hall signatures of this quantum phase transition, which can be described by $N_f$\,$=$\,$2$ QCD$_3$ \cite{ZouandHe2020}. While we initially observe a nonzero $\kappa_{xy}/T$ at zero temperature for $h \lesssim h_{c_2} = 5.5$, this disappears in the partially polarized phase, for which $\lim_{T\rightarrow 0}\, (\kappa_{xy}/T)$\,$=$\,$0$. Intuitively, this trivial response can be deduced from Eq.~\eqref{eq:k_xy0T} as the sum of the Chern numbers of the occupied bands is zero. The trends of $\kappa_{xy}/T$ in Fig.~\ref{fig:th_kitaev}(b) can once again be understood, at least at low temperatures, in terms of the Berry curvatures for the $n$\,$=$\,$2$ and $n$\,$=$\,$3$ bands, plotted in Figs.~\ref{fig:berry}(c) and (d), respectively. Note that the mean-field $h_{c_2}$ does not match the critical field predicted in numerics, which is unsurprising since we neglect the possibility that the magnetic field can also nontrivially renormalize the other couplings in the Hamiltonian. Moreover, while the sum of Chern numbers in the occupied bands is zero in the polarized state, the Chern number of each individual band is nontrivial. However, because we expect the excitations in the polarized state are "spin-flips", which are bound pairs of spinons, the spinon band topology is not directly relevant.

Thus, we have demonstrated the origin of a large but unquantized thermal Hall conductivity in Kitaev materials, such as $\alpha$-RuCl$_3$, without assuming any spin-orbit coupling terms in the Hamiltonian beyond those already in the original Kitaev model \eqref{eq:kitaev}. This is to be contrasted with the scenario proposed by Ref.~\cite{gao2019thermal}, in which the spinons experience an emergent Lorentz force in the applied field due to additional Dzyaloshinskii-Moriya (DM) interactions \cite{dzyaloshinsky1958thermodynamic, moriya1960}. Such a mechanism relies on the field generating a finite second-NN scalar spin chirality on the honeycomb lattice through the DM interaction, thereby inducing an internal gauge flux for the spinons, which gives rise to thermal Hall transport. However, our calculations above show that the unquantized behavior of the thermal Hall effect does not hinge on DM interactions, the microscopic forms of which are presently unclear \cite{winter2016challenges}, but rather, is a much more general phenomenon.

Recall that we can also obtain a nonzero thermal Hall conductivity of $\mc{O}(k_B^2/\hbar)$ for a generic vector $\vec{h}$\,$\|$\,$[11x]$, which lies on the $ac$ plane. A special case of this is, of course, the $[111]$ direction that we have considered so far. To generalize our previous results, we present in Fig.~\ref{fig:th_kitaev}(c) and (d) the temperature dependence of $\kappa_{xy}$ for other polar angles, $\theta$, of the magnetic field  \cite{egmoon2020}, with strengths corresponding to the SFS and ITO phases, respectively. Note that the variation with $\theta$ changes not only the Zeeman term in Eq.~\eqref{eq: mean field Zeeman} but also the coefficient $J_2$\,$\propto$\,$h_xh_yh_z$. 
Interestingly, we observe in Fig.~\ref{fig:k_xy0T}(c) that, for small rotation angles around $\theta = \pi/8$, the zero-temperature value of $\kappa_{xy}/T$ is enhanced from the quantized value in the ITO phase. While $\kappa_{xy}$ is almost invariant under change of sign of the angle $\theta$ in the SFS state (not shown for clarity), there is a clear anisotropy in the ITO phase [see Fig.~\ref{fig:k_xy0T}(d)], as was noticed earlier \cite{yokoi2020half, gordon2020testing}; upon increasing $\theta$ from $-\pi/2$, we see that $\kappa_{xy}/T$ is first half-quantized at a positive value, and subsequently changes sign between $\theta$\,$=$\,$-\pi/4$ and $\theta$\,$=$\,$-\pi/6$ to a negative value.

Finally, in Fig.~\ref{fig:k_xy0T}(e) and (f), we show the magnetic-field dependence of $\kappa_{xy}/T$ at two different temperatures for different angles $\theta$. At a low but finite temperature in Fig.~\ref{fig:k_xy0T}(e), $\kappa_{xy}/T$ decreases smoothly for small magnetic field. As the field's magnitude increases to $|\vec{h}|$ = 0.6, a plateau at $-\pi/12$ for $\theta$\,$=$\,$0$ and $\theta$\,$=$\,$\pi/24$ indicates the ITO phase. We note that for lower temperatures $T/J_1$\,$<$\,$0.1$ (not shown), the plateau persists for a wider range of fields. Further increase of the magnitude of the field to $|\vec{h}|$\,$=$\,$0.8$ shows a phase transition to the SFS phase. At a higher temperature, Fig.~\ref{fig:k_xy0T}(f), the thermal Hall conductivity dies off approaching zero as expected. 

\section{Triangular-lattice Heisenberg antiferromagnet}

The Kitaev model, studied in the previous sections, provided a natural platform to probe the thermal Hall transport in a gapless U(1) spin liquid: associated with the onset of Fermi surfaces, we found an additional zero-temperature contribution to $\kappa_{xy}$ that destroys the original quantization. To gain more insight into this generic behavior, we now turn to a different class of spin models: Heisenberg antiferromagnets. 
Recent numerical evidence  \cite{gong2019chiral} suggests that, on a triangular lattice, the Heisenberg model with competing interactions offers another example of a quantum spin liquid with emergent Fermi surfaces. Importantly for our purposes, the physics  of this system, which is fully spin-rotation invariant, is inherently different from the Kitaev model that, by construction, relies on spin-orbit coupling. 

The Heisenberg model on the triangular lattice has long been the prototype to understand the effects of competing interactions on magnetic orders \cite{kadowaki1990, kimura2006, ye2007, seki2008} and potential QSL states \cite{paddison2017} in several materials. We here consider exchange interactions up to third NNs with the corresponding Hamiltonian
\begin{equation}
\label{eq:heisenberg}
H^{}_{\textsc{h}}=\mc{J}_1\sum_{\braket{i,j}}\Vec{S}_i\cdot\Vec{S}_j+\mc{J}_2\sum_{\braket{\braket{i,j}}}\Vec{S}_i\cdot\Vec{S}_j+\mc{J}_3\sum_{\braket{{\braket{\braket{i,j}}}}}\Vec{S}_i\cdot\Vec{S}_j,
\end{equation}
where $\mc{J}_n >0$ stands for the strength of the $n$-th-NN exchange coupling; in particular, the inclusion of further interactions beyond NNs alone is believed to be an important ingredient in stabilizing QSLs. In such a frustrated system, quantum fluctuations can induce QSL states in the vicinity of classical phase boundaries between different magnetic orders \cite{read1991large, sachdev1992kagome}. 

Even with $\mc{J}_3$\,$=$\,$0$, the model in Eq.~\eqref{eq:heisenberg} is widely recognized to host a spin-liquid phase. The nature of this so-called $\mc{J}_1$-$\mc{J}_2$ spin liquid has been a subject of extensive debate, with several proposed scenarios including a gapless U(1) Dirac spin liquid \cite{kaneko2014gapless, iqbal2016spin, hu2019dirac}, a gapped $\mathbb{Z}_2$ spin liquid \cite{zhu2015spin, zheng2015classification, saadatmand2016}, or competing spin
liquid states \cite{hu2015competing} among others. The story is even richer upon adding the $\mc{J}_3$ coupling; it is believed that the $\mc{J}_1$-$\mc{J}_2$ spin liquid can then extend to a larger parameter range \cite{yao2018quantum, iaconis2018spin}. In a recent work, \citet{gong2019chiral} studied the $\mc{J}_1$-$\mc{J}_2$-$\mc{J}_3$ Heisenberg model using the DMRG algorithm. Choosing $\mc{J}_1 = 1.0$ as the overall energy scale, they identified a CSL phase in the coupling range $0 \leq \mc{J}_2/\mc{J}_1 \leq 0.7, 0 \leq \mc{J}_3/\mc{J}_1 \leq 0.4$, in proximity to the previously found $\mc{J}_1$-$\mc{J}_2$ spin liquid and the triple point of the different magnetic orders. Unlike the gapped CSL phase on the kagom\'{e} lattice \cite{he2014chiral, gong2014, gong2015global}, the CSL phase harbored by the triangular lattice is gapless. This state spontaneously breaks time-reversal symmetry with a finite scalar chiral order $\langle\vec{S}_i\cdot(\vec{S}_j \times \vec{S}_k) \rangle$ for the three spins $i,j,k$ on a triangular plaquette.
Moreover, the large central charge estimated numerically is redolent of a scenario with emergent spinon Fermi surfaces \cite{ioffe1989gapless, nagaosa1990normal, motrunich2005variational,sheng2009spin}. In light of these observations, we will now try to understand the thermal Hall effect in this CSL phase and compare the response to that previously evaluated for the Kitaev model.

\begin{figure}[tb]
\includegraphics[width=0.9\linewidth]{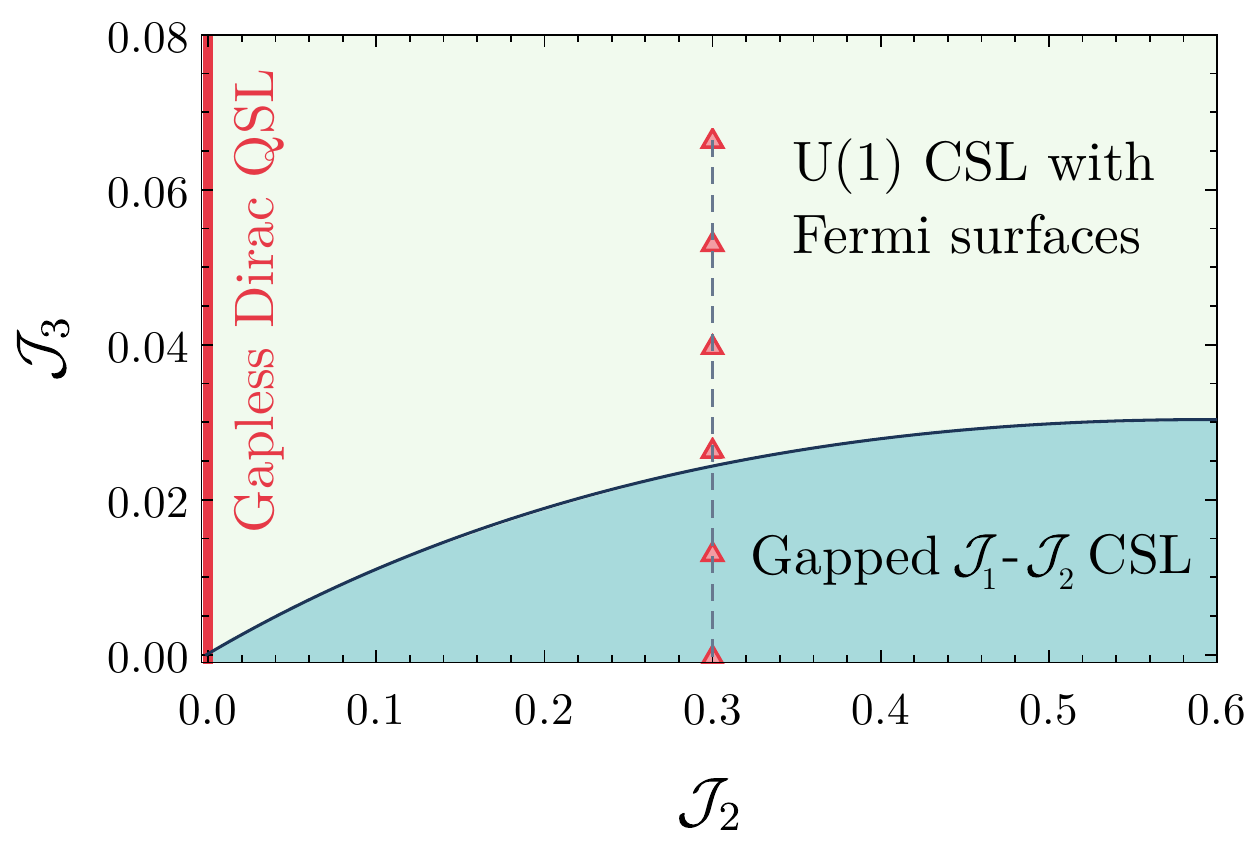}
\caption{Phase diagram of the $\mc{J}_1$-$\mc{J}_2$-$\mc{J}_3$ Heisenberg model in mean-field theory [see Eq.~\eqref{eq:hH-MF}]. At $\mc{J}_2$\,$=$\,$0$, the spectrum is gapless and hosts a pair of Dirac cones. By varying $\mc{J}_3$ while keeping $\mc{J}_2$ fixed, the system can be driven from a gapped $\mc{J}_1$-$\mc{J}_2$ spin liquid, with a quantized zero-temperature $\kappa_{xy}/T$ to a gapless U(1) CSL with emergent spinon Fermi surfaces and an unquantized response. The red points indicate the values of $\mc{J}_3$ considered for computing the thermal Hall conductivity in Fig.~\ref{fig:th_H}(a).}
\label{fig:pd}
\end{figure}

\subsection{Model}

In order to explain the abovementioned DMRG results, Ref.~\cite{gong2019chiral} proposed a staggered flux state, which could account for both the Fermi surfaces in the gapless CSL and the observed coexisting chiral edge modes \cite{li2008entanglement, dubail2015tensor}. As before, we construct a mean-field theory for this state using the Abrikosov-fermion representation of spin-$1/2$ operators \eqref{eq: parton-usual-2}. Due to the SU(2) spin-rotational symmetry, the Hamiltonian simplifies considerably using the identity
\begin{equation*}
\hat{S}^x_i \hat{S}^x_j + \hat{S}^y_i \hat{S}^y_j + \hat{S}^z_i \hat{S}^z_j
= - \frac{1}{2} c^\dag_{i \alpha} c^{\pdagger}_{j \alpha} c^\dag_{j \beta} c^{\pdagger}_{i \beta}
+ \frac{1}{4} c^\dag_{i \alpha} c^{\pdagger}_{i \alpha} c^\dag_{j \beta} c^{\pdagger}_{j \beta},
\end{equation*}
with repeated indices implicitly summed over.
In a U$(1)$ QSL, all spinon pairing terms must necessarily vanish. Thus, carrying out the mean-field decoupling with the assumption that only fermionic hopping terms acquire nonzero expectation values, the Heisenberg Hamiltonian $H \simeq \sum_{{i,j}}\mc{J}_{ij}\,\mathbf{S}_{i}\cdot\mathbf{S}_{j}$ reads (up to constants) as
\begin{alignat}{1}
    H_{\textsc{h}}^{\textsc{mf}} &=\sum_{\langle{ij}\rangle} \frac{\mc{J}^\pdagger_{ij}}{4}\sum_{\alpha}\left(-{\zeta}^{*}_{ij}\,c_{i, \alpha}^{\dagger}c^{\pdagger}_{j, \alpha}+\mathrm{h.c.}\right)+\sum_{\langle{ij}\rangle}\frac{\mc{J}^\pdagger_{ij}}{4} \,\lvert\zeta^{\pdagger}_{ij}\rvert^{2}; \nonumber\\
\zeta^{\pdagger}_{ij} &\equiv \sum_{\alpha}\langle c_{i, \alpha}^{\dagger}c^\pdagger_{j, \alpha} \rangle = {\zeta}^{*}_{ji}.
\label{eq:HHMF}
\end{alignat}
Note that in deriving Eq.~\eqref{eq:HHMF}, we have made use of the single-occupancy constraint \eqref{eq: gauge constraint} on the parton Hilbert space to eliminate on-site terms such as $\langle c^\dagger_{i\alpha} c^\pdagger_{i\alpha} \rangle$ at the mean-field level. 

In principle, the $\zeta_{ij}$ can be solved for self-consistently but here, for the sake of generality, we treat them as free (bounded) parameters. The expectation values $\{\zeta_{ij}\}$ then collectively define a mean-field \textit{ansatz}. The projective action of lattice or time-reversal symmetries on this ansatz describes the particular spin-liquid state of interest. Specifically, we focus on a U$(1)$ spin liquid known as the staggered flux state~\cite{wen2002quantum, bieri2016projective, li2017spinon}. Its PSG specifies that the fermionic spinons transform as 
\begin{equation}
c^\pdagger_\alpha\,(\vec{r}) \xrightarrow{\mc{T}_{1}}c^\pdagger_\alpha\,(\vec{r}+{\bf a}^{}_1),\,\,
  c^\pdagger_\alpha\,(\vec{r}) \xrightarrow{\mc{T}_{2}}(-)^{r_{1}}c^\dagger_\alpha\,(\vec{r}+{\bf a}^{}_2),
    \label{eq:mag_trans}
\end{equation}
under translations $\mc{T}_{1}$ and $\mc{T}_{2}$ along the unit vectors ${\bf a}_1$\,$=$\,$(1, 0)$ and ${\bf a}_{2}$\,$=$\,$(1, \sqrt{3})$ of the triangular lattice, respectively. As can be seen from the factor of $(-1)^{r_{1}}$ in Eq.~\eqref{eq:mag_trans}, the mean-field ansatz is translationally invariant only modulo a gauge transformation, thus necessitating the use of a two-site unit cell when working in a fixed gauge. In spite of the unit cell being doubled, the projected wavefunction, of course, preserves the lattice translation symmetries along both ${\bf a}_{1,2}$.

\begin{figure}[tb]
\includegraphics[width=\linewidth]{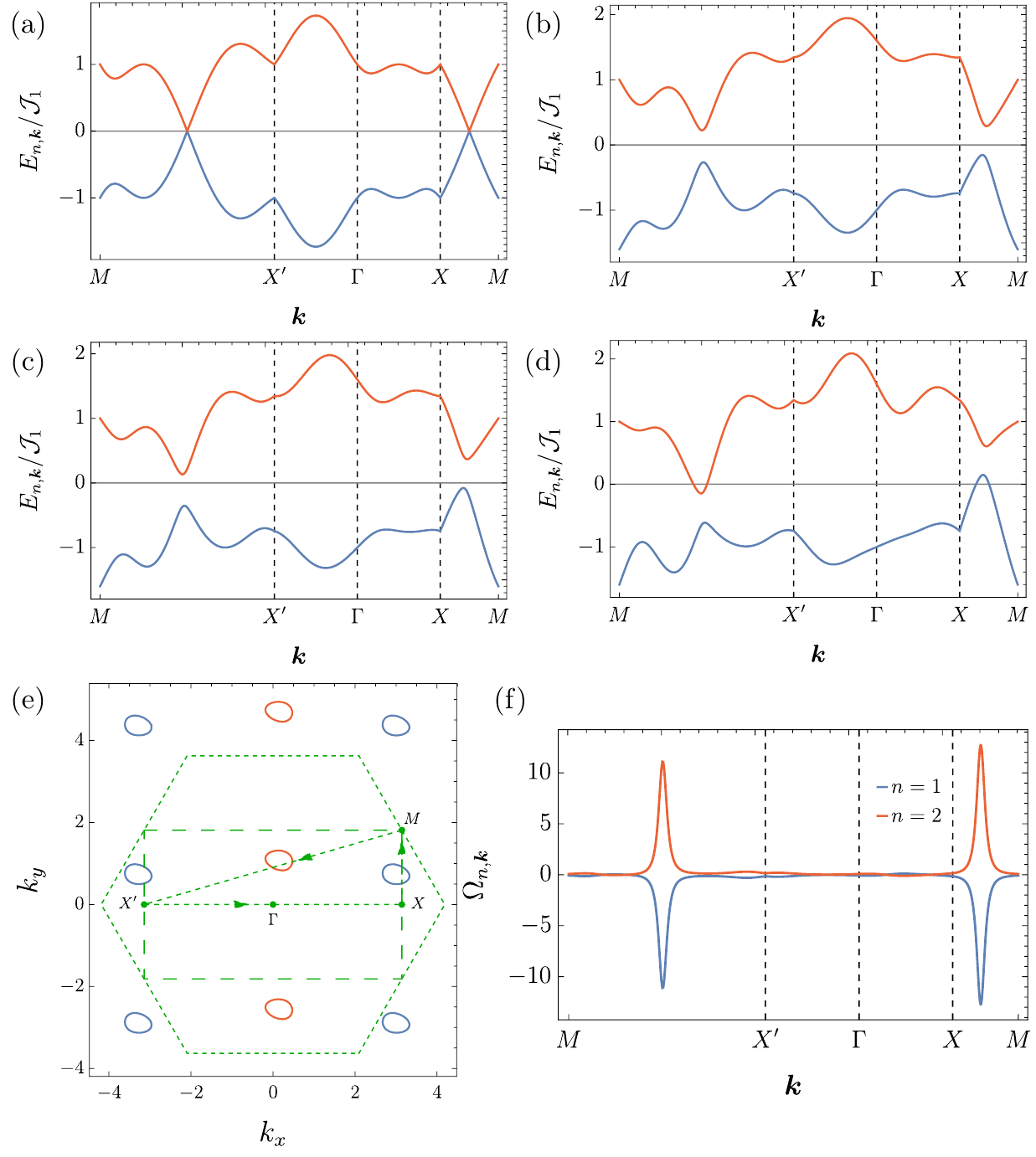}
\caption{Band structure of the staggered-flux ansatz [Eq.~\eqref{eq:hH-MF}] along high-symmetry lines with mean-field parameters as follows. The amplitudes of the first-, second-, and third-NN hoppings are $\zeta$\,$=$\,$1.0$, $\lambda$\,$=$\,$1.0$, and $\rho$\,$=$\,$3.0$; the associated phases are $\phi_{1}$\,$=$\,$\phi_{2}$\,$=$\,$\pi/2$, $\varphi_{1}$\,$=$\,$\varphi_{2}$\,$=$\,$\varphi_{3}$\,$=$\,$0$, and $\gamma_{1}$\,$=$\,$\gamma_{2}$\,$=$\,$\gamma_{3}$\,$=$\,$\pi/2$, respectively. Each band is doubly degenerate as the two spinon species have identical energies. (a) At $\mc{J}_2$\,$=$\,$\mc{J}_3$\,$=0$, there are two Dirac cones in the spectrum. (b) Adding a nonzero $\mc{J}_2$\,$=$\,$0.3$ gaps them out. (c) A small $\mc{J}_3$\,$=$\,$0.04/\rho$ shifts the two cones unequally; note that the state is still gapped at this stage. (d) Finally, beyond a threshold $\mc{J}_3$, Fermi surfaces emerge as plotted here for $\mc{J}_3$\,$=$\,$0.16/\rho$. (e) The blue and yellow pockets trace out electron-like and hole-like Fermi surfaces, respectively. Owing to our choice of a two-site unit cell, the Brillouin zone is defined as the region enclosed by the dashed rectangle. (f) The Berry curvature is a function of $\mc{J}_2$ alone; for each of the two bands, its distribution in momentum space is peaked at the $\vec{k}$-vectors of the original Dirac points.}
\label{fig:h-bands}
\end{figure}

The mean-field phase diagram determined from this ansatz is sketched in Fig.~\ref{fig:pd}.
While the explicit form of the Hamiltonian is detailed in Appendix~\ref{app:staggered}, let us briefly discuss its structure here. 
Since the underlying spin model retains couplings up to third-NN sites, it is only natural to allow for all symmetry-allowed hopping terms up to the same range in the mean-field ansatz.  To wit, we take
\begin{equation}
\label{eq:hH-MF}
H_{\textsc{h}}^{\textsc{mf}}=\mc{J}^{}_{1}\mc{H}^{}_{1}+\mc{J}^{}_{2}\mc{H}^{}_{2}+\mc{J}^{}_{3}\mc{H}^{}_{3},
\end{equation}
where $\mc{H}_i$ describes hopping between $i$-th neighbors.
The first-NN hopping processes are characterized by an amplitude $\zeta$ and two phases $\phi_1, \phi_2$. On setting $\phi_1 $\,$=$\,$ \phi_2 $\,$=$\,$ \pi/2$, the ansatz with NN hopping alone reduces to the familiar $\pi$-flux U$(1)$ QSL state \cite{lu2016symmetric}, and the spectrum exhibits a pair of Dirac cones, centered at half filling, for each spinon species as drawn in Fig.~\ref{fig:h-bands}(a). However, one can also engineer any other value of the flux threading each plaquette through a suitable choice of the phases $\phi_{1,2}$.
Inclusion of the second- (or third-) NN hoppings result in the opening of a direct gap at each Dirac cone. The resultant (fully gapped) bands are topologically nontrivial, with Chern numbers $C$\,$=$\,$\pm1$; from the bulk-boundary correspondence \cite{thouless1982quantized, hatsugai1993chern}, this gives rise to the chiral edge state. Lastly, the third-NN hoppings split the degeneracy of the two Dirac cones, generating a particle-like and a hole-like SFS, one around each Dirac point [blue and red, respectively, in Figs.~\ref{fig:h-bands}(e)]. These terms do not alter any of the topological properties, which are controlled instead by $\mc{J}_2$. Since the mean-field ground state is, once again, at half-filling by virtue of Eq.~\eqref{eq: gauge constraint}, the particle- and hole-like Fermi surfaces are always perfectly compensated. 

\begin{figure*}[htb]
\includegraphics[width=\linewidth]{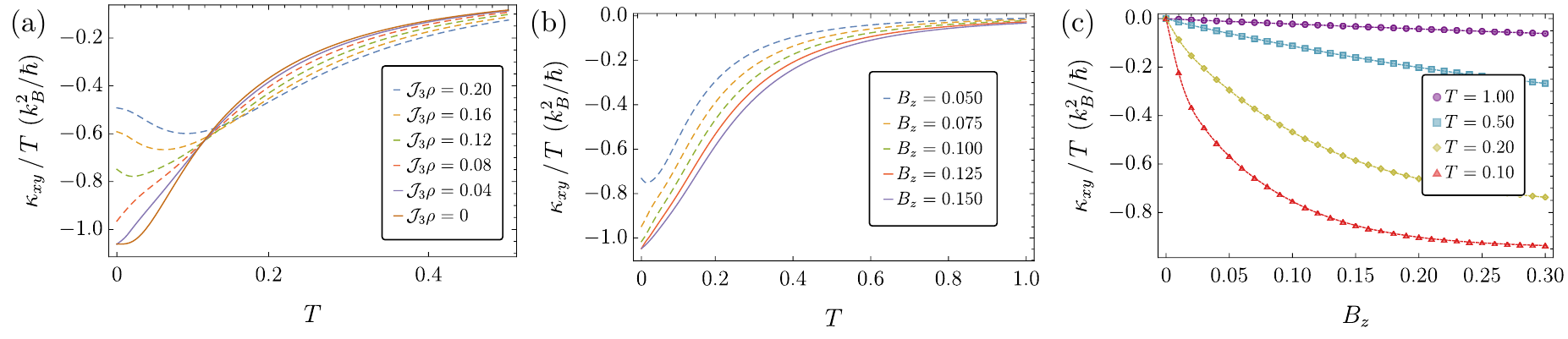}
\caption{Temperature dependence of the thermal Hall conductivity for the $\mc{J}_1$-$\mc{J}_2$-$\mc{J}_3$ Heisenberg model described by the mean-field theories (a) $H_{\textsc{h}}^{\textsc{mf}}$ [Eq.~\eqref{eq:hH-MF}], and (b) $\widetilde{H}_{\textsc{h}}^{\textsc{mf}}$ [Eq.~\eqref{eq:hH-MF2}]. In both cases, the system can be tuned from the $\mc{J}_1$-$\mc{J}_2$ spin liquid, with a quantized thermal Hall coefficient of $\pi/3$, to a CSL, endowed with spinon Fermi surfaces and an unquantized response, by varying either (a) the third-NN coupling $\mc{J}_3$, or (b) the magnetic field $B_z$. The precise mean-field parameters used for the computation of $\kappa_{xy}/T$ are detailed in Figs.~\ref{fig:h-bands}(b)-(d) and \ref{fig:ansatz}, respectively. (c) Field dependence of $\kappa_{xy}/T$ of $\widetilde{H}_{\textsc{h}}^{\textsc{mf}}$ for a magnetic field in the $z$ direction. As expected, the thermal Hall conductivity dies off at higher temperatures.}
\label{fig:th_H}
\end{figure*}

So far, we have regarded the couplings $\mc{J}_i$ as material parameters which are inherent to the particular quantum magnet under consideration and thus, cannot be easily varied. For the purpose of tunability, therefore, it is useful to consider the coupling to an external field, as a function of which, the system can be driven across a phase transition between the $\mc{J}_1$-$\mc{J}_2$ spin liquid and the CSL. Note that time-reversal symmetry is already broken by a nonzero $\mc{J}_2$ in Eq.~\eqref{eq:hH-MF} (i.e., there does not exist a gauge transformation which, combined with the action of the TRS operation, leaves the ansatz invariant)---this points toward a natural route to incorporating the effect of a magnetic field, as in the Kitaev model. However, $\mc{H}_2$ also breaks further symmetries, including reflections, $C_2$ rotation, as well as the combination of reflection and time-reversal: so, as it stands, $\mc{H}_2$ lacks the correct symmetries to describe a physical magnetic field perpendicular to the plane of the system. Similar considerations apply to $\mc{H}_3$ as well; in fact, it can be shown for this model that any term breaking the degeneracy of the Dirac points cannot have the symmetries of a perpendicular magnetic field (see Appendix \ref{app:staggered}).

To correct for this shortcoming, we now construct an alternative mean-field Hamiltonian which resembles Eq.~\eqref{eq:hH-MF} but with the minimal modification that the second-NN hopping term mimics the orbital coupling of a magnetic field \cite{sen1995large} from the point of view of symmetries. The spinons transform as
\begin{subequations}
\begin{alignat}{2}
c_\alpha (\vec{r})&\xrightarrow[]{C_2} \, &&(-)^{r_1+r_2}\,c_\alpha\left(C^{}_2 \vec{r}\right),\\
c_\alpha(\vec{r}) &\xrightarrow[]{\Theta}\, &&(-)^{r_1+r_2}\,c_{\bar{\alpha}}(\vec{r}),\\
c_\alpha(\vec{r}) &\xrightarrow[]{R_1}\, &&\mathrm{e}^{-i\,r_2\pi/2}\,c_\alpha\left(R^{}_1 \vec{r}\right),\\
c_\alpha(\vec{r}) &\xrightarrow[]{R_2}\, &&\mathrm{e}^{i\,(r_1\pi-r_2\pi/2)}c_\alpha\left(R^{}_2 \vec{r}\right),
\end{alignat} 
\end{subequations}
where $R_1$($R_2$) refers to the horizontal (vertical) axis of reflection. The modified term, which we label $\widetilde{\mc{H}}^{}_{2}$, breaks time-reversal and both reflections, but preserves $C_2$ and $\Theta\,R_i$. Including a coupling to a Zeeman field in the $\hat{z}$ direction as well, the combined Hamiltonian, shown in Appendix~\ref{app:staggered}, assumes the form
\begin{equation}
\label{eq:hH-MF2}
\widetilde{H}_{\textsc{h}}^{\textsc{mf}}=\mc{J}^{}_{1}\mc{H}^{}_{1}+\widetilde{\mc{J}}^{}_{2} \widetilde{\mc{H}}^{}_{2}+\mc{J}^{}_{3}\mc{H}^{}_{3} - \frac{1}{2} \sum_{i,\alpha} (-)^\alpha  B^{}_{z}\, c^\dagger_{i, \alpha} c^\pdagger_{i, \alpha},
\end{equation}
with $\widetilde{\mc{J}}_2$\,$\propto$\,$B_{z}$.
The physics of this model, as we will see below, is similar to that of Eq.~\eqref{eq:hH-MF}.

\subsection{Thermal Hall conductivity}
\label{sec:th_h}

Since the projective actions of translation symmetries in Eq.~\eqref{eq:mag_trans} dictate the use of a two-sublattice unit cell, the Fourier-transformed counterparts of both $H_{\textsc{h}}^{\textsc{mf}}$ and  $\widetilde{H}_{\textsc{h}}^{\textsc{mf}}$ can be compactly expressed in momentum space by using a spinor structure identical to Eq.~\eqref{eq:Hk1}. This implies that four bands are obtained upon diagonalization of the respective kernels. With regard to $H_{\textsc{h}}^{\textsc{mf}}$, the bands for the different spinon species are degenerate, as conveyed by Fig.~\ref{fig:h-bands}. This degeneracy is split by the Zeeman field in $\widetilde{H}_{\textsc{h}}^{\textsc{mf}}$; the dispersions of this model for relevant parameters are arrayed in Fig.~\ref{fig:ansatz} of Appendix~\ref{app:staggered}. 

Armed with the band structures, the thermal Hall conductivities of our two mean-field theories for the $\mc{J}_1$-$\mc{J}_2$-$\mc{J}_3$ Heisenberg model can now be computed using Eq.~\eqref{eq:k_xy}; the final results are illustrated in Fig.~\ref{fig:th_H}. Beginning with $H_{\textsc{h}}^{\textsc{mf}}$, at small $\mc{J}_3$\,$\ll$\,$\mc{J}_1$, we notice that $\kappa_{xy}/T$  is quantized for $T$\,$\rightarrow$\,$0$ [Fig.~\ref{fig:th_H}(a)], as before. However, the key difference with Fig.~\ref{fig:th_kitaev} lies in that the plateau occurs at $\pi/3$ (gauge fluctuations modify this to $\pi/6$ \cite{guo2020gauge}), as opposed to $-\pi/12$ for the Kitaev model. Recognizing that the Chern number of the lower band in Fig.~\ref{fig:h-bands}(c) is $-1$, this fourfold-enhanced transport coefficient can be explained straightforwardly from Eq.~\eqref{eq:k_xy0T}, multiplied by an additional factor of 2 to account for the two spinon species. As the Fermi surfaces develop for larger $\mc{J}_3$, the zero-temperature value strays from the quantized number; following arguments analogous to Sec.~\ref{sec:results_kitaev}, the sign of this deviation can be intuited by inspecting the profiles of the Berry curvatures in the Brillouin zone [Fig.~\ref{fig:h-bands}(f)]. 
Finally, we also study the thermal Hall response of $\widetilde{H}_{\textsc{h}}^{\textsc{mf}}$ in Fig.~\ref{fig:th_H}(b), as a function of the magnetic field $B_{z}$, which takes the system from a fully gapped phase to one with emergent Fermi surfaces. The general features of the conductivity are comparable to those seen in Fig.~\ref{fig:th_H}(a), in terms of both the quantization (or lack thereof) as well as the overall magnitude. It is interesting to note, however, that for any given $T$, $\kappa_{xy}/T$ is always monotonically increasing with $B_z$; such uniform monotonicity is absent in the case of $H_{\textsc{h}}^{\textsc{mf}}$, for which $\kappa_{xy}/T$ can either increase or decrease as $\mc{J}_3$ is tuned, depending on the temperature range. This effect is also shown in Fig.~\ref{fig:th_H}(c), which illustrates the thermal Hall response as a function of the field magnitude $B_z$.

\section{Summary and conclusion}
In this work, we analyzed the thermal Hall conductivity in two important models of quantum magnets---the Kitaev honeycomb lattice model and the $\mc{J}_1$-$\mc{J}_2$-$\mc{J}_3$ Heisenberg magnet on the triangular lattice. We paid special attention to the impact of the magnetic field, taking into account that it can drive the gapped QSL phases these systems harbor into gapless QSLs with spinon Fermi surfaces, as indicated by recent numerical studies \cite{zhu2018robust, liang2018intermediate, gohlke2018dynamical,nasu2018successive,hickey2019emergence,ronquillo2019signatures,patel2019magnetic,gong2019chiral}. For our computations, we employed a mean-field description of these phases, based on fermionic spinons, that is constrained by the aforementioned numerics and a PSG analysis.

For the Kitaev honeycomb lattice model in a magnetic field, $H^{}_{\textsc{k}} + H^{}_\textsc{z}$ defined in Eqs.~(\ref{eq:kitaev}) and (\ref{eq:Bfield}), our analysis captures three phases: as illustrated in Fig.~\ref{fig:phases}, the non-Abelian ITO phase, which emerges when gapping out the Kitaev $B$ phase by a small magnetic field, transitions into a gapless U(1) QSL at an intermediate value, $h_{c1}$, of the magnetic field. In the fermionic spinon language, this corresponds to the magnetic-field-induced loss of $p_x + i p_y$ superconducting pairing; we capture this by the mean-field Hamiltonian in Eq.~(\ref{HamiltonianBothPhases}) where the $\xi(\vec{h})$ describes the vanishing of the superconducting term. Besides $\xi(\vec{h})$, the magnetic field also enters as the usual Zeeman term and nonlinearly induces a second-NN hopping $J_2 \propto h_x h_y h_z$ of the spinons. As a result of the finite gap, the thermal Hall conductivity $\kappa_{xy}/T$ is quantized at zero temperature in the ITO phase, see solid lines in Fig.~\ref{fig:k_xy0T}(a); it reaches $-\pi/12$ in units of $k_B^2/\hbar$ at $T=0$, resulting from the Chern numbers of the BdG spinon bands, and is associated with the presence of Majorana edge modes. When the superconducting pairing disappears at $h_{c1}$, we obtain spinon Fermi surfaces and $\lim_{T \rightarrow 0} \kappa_{xy}/T$ is not quantized any more, but varies continuously with magnetic field, see dashed lines in Fig.~\ref{fig:k_xy0T}(a) and (b). We have related its increase with $|\vec{h}|$ to the distribution of the Berry curvature and Fermi surfaces of the spinons. 
For larger magnetic fields, we eventually reach a phase at $|\vec{h}|=h_{c2}$ that is adiabatically connected to the fully polarized state. This corresponds to a gapped spinon band structure with vanishing net Chern number in the occupied bands and associated vanishing $\kappa_{xy}/T$ at zero temperature, see solid lines in Fig.~\ref{fig:k_xy0T}(b). We have also studied in detail the predicted dependence as a function of the direction of the magnetic field, as summarized in Fig.~\ref{fig:k_xy0T}(c)-(f), both for the ITO and SFS phases. 

In the second part of the paper, we have performed a similar analysis for the triangular-lattice Heisenberg model with exchange interactions up to third NNs, which is experimentally relevant as a low-energy description of various QSL candidate materials \cite{shimizu2003, kurosaki2005, yamashita2008, yamashita2009, yamashita2010,Law6996, yu2017, ribak2017, klanjsek2017,kadowaki1990, kimura2006, ye2007, seki2008, daidai2020,li2015, shen2016evidence,paddison2017,li2017,paddison2017, zhu2018, zhang2018, shen2018}. Motivated by a recent numerical study \cite{gong2019chiral} indicating that this model can host a gapless spin-liquid phase with nonvanishing chiral spin correlations, we study two different ans\"atze, Eqs.~(\ref{eq:hH-MF}) and (\ref{eq:hH-MF2}), that can capture the transition from a gapped to a gapless CSL on the triangular lattice, see Fig.~\ref{fig:pd}. Unlike the Kitaev honeycomb-lattice model, this model does not involve any spin-orbit coupling; nonetheless, the behavior of $\kappa_{xy}/T$ is qualitatively similar, as can be seen in Fig.~\ref{fig:th_H}: in the gapped phase [solid lines in Fig.~\ref{fig:th_H}(a) and (b)], $\kappa_{xy}/T$ is quantized as $T\rightarrow 0$, albeit with a value four times larger, resulting from the presence of spinful and complex spinons (as opposed to the nondegenerate bands of Majorana fermions in our description of the ITO phase of the Kitaev model); in the gapless phase with a spinon Fermi surface, we again observe that the low-temperature thermal Hall conductivity is not quantized and reduced in magnitude (see dashed lines).  

Taken together, our analysis shows that a proper description of the thermal Hall conductivity in a QSL requires taking into account the effect of the magnetic field on the parameters of the underlying parton ansatz. We believe that a detailed comparison of our predictions with future measurements of $\kappa_{xy}$ will help shed light on the possible QSL phases hosted by ``Kitaev materials'' and other frustrated magnets. 

Finally, we note that our computations were in the context of a spinon mean-field theory. This yields the correct exact value of the thermal Hall conductivity in the non-Abelian ITO phase. Gauge fluctuations will be important in the other phases, and some discussion of their consequences appears elsewhere \cite{guo2020gauge}.



\subsection*{Acknowledgements}

We thank Gang Chen, G. Grissonnanche, and L. Taillefer for stimulating discussions.
This research was supported by the National Science Foundation under Grant No.~DMR-2002850.

\appendix

\section{Mean-field theory for the Kitaev model}
\label{app:kitaev_f_H}

In this appendix, we discuss the representation of the Kitaev model \cite{kitaev2006anyons} in terms of fermionic spinons, following closely the analyses of Refs.~\cite{you2012doping} and \cite{ZouandHe2020}. Each two-spin term in the original model \eqref{eq:kitaev} can be written as a product of four Majorana fermions, as outlined in Eq.~\eqref{eq:zero field}. Carrying out a systematic mean-field decoupling, the Kitaev model reduces to a \textit{quadratic} Hamiltonian of Majorana fermions $\chi_{A,B}^{\alpha}$, with $\alpha$\,$=$\,$0,x,y,z$, where $A (B)$ connotes the sublattice index.  This consists of three types of terms:
\begin{equation} \label{eq: Kitaev-original}
H^{\textsc{mf}}_{\textsc{k}}=H^{}_1+H^{}_2+H^{}_3
\end{equation}
where $H_{1,3}$ involve NN couplings while $H_2$ couples next-NNs. These are given by
\begin{widetext}
\begin{equation}
\label{eq:maj_k}
\begin{split}
	H^{}_1=2iJ_1\sum_{\vec r_i}\Bigg(&\chi_B^0(\vec r_i)\chi_A^0(\vec r_i)+\chi_B^0(\vec r_i+\vec n_2)\chi_A^0(\vec r_i)
	+\chi_B^0(\vec r_i-\vec n_1)\chi_A^0(\vec r_i)\Bigg),\\
	H^{}_2=2iJ_2\sum_{\vec r_i}\Bigg(
	&\chi^0_A(\vec r_i+\vec n_1)\chi^0_A(\vec r_i) +\chi^0_A(\vec r_i+\vec n_2)\chi^0_A(\vec r_i)
	+\chi^0_A(\vec r_i+\vec n_3)\chi^0_A(\vec r_i)\\
	-&\chi^0_B(\vec r_i+\vec n_1)\chi^0_B(\vec r_i)
	-\chi^0_B(\vec r_i+\vec n_2)\chi^0_B(\vec r_i) -\chi^0_B(\vec r_i+\vec n_3)\chi^0_B(\vec r_i\Bigg),\\
	H^{}_3=2iJ_1'\sum_{\vec r_i}\Bigg(&\chi^z_A(\vec r_i)\chi^z_B(\vec r_i)+\chi^x_A(\vec r_i-\vec n_2)\chi_B^x(\vec r_i)
	+\chi_A^y(\vec r_i+\vec n_1)\chi_B^y(\vec r_i)\Bigg),
\end{split}
\end{equation}
\end{widetext}
with $\vec{n}_{1,2}$ denoting the lattice vectors along the directions corresponding to  translations $T_{1,2}$ in Fig.~\ref{fig:symmetry}(a).  Our focus will be not so much on the precise values of $J_1$, $J_2$ and $J_1'$ (which can, in principle, be solved for self-consistently) but rather, on the set of phases that can be obtained by varying them as free parameters.

Next, we need to specify how the Abrikosov fermions are constructed from the Majoranas but this mapping is certainly not unique. 
\citet{you2012doping} relate the spinon $c$, defined in Eq.~\eqref{SpinonMatrix}, to the Majorana fermions via
\begin{equation}
c^{}_1=\frac{1}{\sqrt{2}}\left(\chi^0+i\chi^z\right),
\quad
c^{}_2=\frac{1}{\sqrt{2}}\left(i\chi^x-\chi^y\right).
\end{equation}
However, the SU$(2)$ gauge redundancy \eqref{eq: SU(2)-redundancy} connotes that we have the freedom to define another (equally valid) set of partons 
$f_{i,\eta}$, $\eta=1,2$, which are related to (\ref{eq: parton-usual-2}) by a gauge transformation
\begin{equation} \label{eq: parton-transformed}
F_i=
\left(
\begin{array}{cc}
	f_{i,1} & - f_{i,2}^\dag\\
	f_{i,2} & f_{i,1}^\dag
\end{array}
\right)
= \mc{C}_i\, \mc{W}_i,
\end{equation}
where we take $\mc{W}$ to be $\mc{W}_{A (B)}$ on the $A$\,$(B)$ sublattice, such that
\begin{equation} \label{eq: parton-transforming}
\mc{W}^{}_A=a+i(b\,\sigma_1+a\,\sigma_3), \quad
\mc{W}^{}_B=\mathrm{e}^{-i\sigma_3\frac{\pi}{4}}\,\mc{W}_A^*\,\mathrm{e}^{i\sigma_3\frac{\pi}{4}}
\end{equation}
with
\begin{equation}
a=\sqrt{\frac{1}{6-2\sqrt{3}}},\quad\ b=(\sqrt{3}-1)\,a.
\end{equation}
The utility of this exercise lies in that the gauge charge of these new spinons is always preserved under the symmetry transformations. 
The Majorana Hamiltonian \eqref{eq:maj_k} can easily be re-expressed using these modified spinons. In the spirit of Eq.~\eqref{eq: Kitaev parton}, we now decompose each of the three pieces in \eqref{eq: Kitaev-original} individually into hopping and pairing terms, in accordance with the prescription of \citet{ZouandHe2020}. In the following, our equations employ the convention that the first lines on the right-hand-side always contribute to $H_{\rm hopping}$ only, while the second lines add to $H_{\rm pairing}$ alone.

\begin{widetext}

When expanded using the transformed spinons, $H_1$ reads as
\begin{equation}
\begin{split}
	2iJ_1\chi_B^0(\vec r_j)\chi_A^0(\vec r_i)
	=J_1\Bigg\{&f_B^\dag(\vec r_j)\left[\left(a^2+\frac{b^2}{2}\right)+ (a^2-\frac{b^2}{2})\sigma_3+ab\sigma_1+ab\sigma_2\right]f^\pdagger_A(\vec r_i)+\mbox{h.c.}\\
	+&f_B^T(\vec r_j)\left[\left(ia^2+\frac{b^2}{2}\right)+\left(ia^2-\frac{b^2}{2}\right)\sigma_3+ ab(1+i)\sigma_1\right]f^\pdagger_A(\vec r_i)+\mbox{h.c.}\Bigg\}.
\end{split}
\end{equation}
All terms in $H_2$ are of the form $2iJ_2\tau_z\chi^0(\vec r_j)\chi^0(\vec r_i)$, with $\tau_z= + 1$ ($\tau_z=-1$) for the $A$\,$(B)$ sublattice:
\begin{equation}
\begin{split}
	2iJ_2\tau_z\chi^0(\vec r_j)\chi^0(\vec r_i)
	=J_2\Bigg\{&f^\dagger(\vec r_j)\left[i\left(a^2+\frac{b^2}{2}\right)+i\left(a^2-\frac{b^2}{2}\right)\sigma_3 +iab\sigma_1+iab\sigma_2\right]\tau_zf(\vec r_i)+\mbox{h.c.}\\
	+&f^T(\vec r_j)\left[\left(a^2-\frac{ib^2}{2}\right)+\left(a^2+\frac{ib^2}{2}\right)\sigma_3 +ab(1-i)\sigma_1\right]f(\vec r_i)+\mbox{h.c.}\Bigg\}.
\end{split}
\end{equation}
The minus sign between the two sublattices is due to the directed nature of the Majorana hopping stemming from the effective Hamiltonian \eqref{eq:three-spin}; for details of the derivation, we refer the interested reader to Eq.~(48) of Ref.~\cite{kitaev2006anyons}. Finally, the terms in $H_3$, which depend on the type of the bond ($x$, $y$, or $z$), have the spinon representation
\begin{alignat}{3}
\nonumber
	2iJ_1'\chi^x_A(\vec r_j)\chi^x_B(\vec r_i) &=
	J_1'&&\Bigg\{&&f_A^\dag(\vec r_j)\left[\left(-a^2-\frac{b^2}{2}\right)+\left(a^2-\frac{b^2}{2}\right)\sigma_3 -ab\sigma_1+ab\sigma_2\right]f^\pdagger_B(\vec r_i)+\mbox{h.c.}\\
	& &&+&&f_A^T(\vec r_j)\left[\left(-ia^2-\frac{b^2}{2}\right)-\left(-ia^2+\frac{b^2}{2}\right)\sigma_3- ab(1+i)\sigma_1\right]f^\pdagger_B(\vec r_i)+\mbox{h.c.}\Bigg\},\\
	\nonumber
	2iJ_1'\chi_A^y(\vec r_j)\chi_B^y(\vec r_i) &=
	J_1'&&\Bigg\{&&f_A^\dag(\vec r_j)\left[\left(-a^2-\frac{b^2}{2}\right)+\left(a^2-\frac{b^2}{2}\right)\sigma_3 +ab\sigma_1-ab\sigma_2\right]f^\pdagger_B(\vec r_i)+\mbox{h.c.}\\
&	&&+&&f_A^T(\vec r_j)\left[\left(ia^2+\frac{b^2}{2}\right)+\left(-ia^2+\frac{b^2}{2}\right)\sigma_3- ab(1+i)\sigma_1\right]f^\pdagger_B(\vec r_i)+\mbox{h.c.}\Bigg\},\\
\nonumber
	2iJ_1'\chi^z_A(\vec r_j)\chi^z_B(\vec r_i)&=
	J_1'&&\Bigg\{&&f^\dag(\vec r_j)\left[\left(-a^2-\frac{b^2}{2}\right)-\left(a^2-\frac{b^2}{2}\right)\sigma_3 +ab\sigma_1+ab\sigma_2\right]f(\vec r_i)+\mbox{h.c.}\\
&	&&+&&f^T(\vec r_j)\left[\left(-ia^2-\frac{b^2}{2}\right)+\left(-ia^2+\frac{b^2}{2}\right)\sigma_3+ ab(1+i)\sigma_1\right]f(\vec r_i)+\mbox{h.c.}\Bigg\}.
\end{alignat}
\end{widetext}

\section{Ansatz for the staggered flux state}
\label{app:staggered}

In this section, we summarize the details for the staggered flux state proposed by \citet{gong2019chiral} to explain their numerical observation of a gapless CSL in the $\mc{J}_1$- $\mc{J}_2$- $\mc{J}_3$ Heisenberg model. 

The structure of the mean-field theory that we study is specified by Eq.~\eqref{eq:hH-MF}, wherein $\mc{H}_i$ encompasses $i$-th NN hopping processes. The ground state is always at half-filling of the fermionic spinons. For the purpose of the following discussion, it suffices to consider only one of the two species of spinons; the Hamiltonian for the other species is identical.

To begin, we note that the unit cell is doubled in the mean-field ansatz \cite{ying2007}. Defining the three NN vectors
\begin{equation}
\begin{aligned}
    \delta_{1}&=(1, 0) = {\bf a}_1, \\
    \delta_{2}&=\left(\frac{1}{2}, \frac{\sqrt{3}}{2}\right) = \frac{{\bf a}_2}{2}, \\
    \delta_{3}&=\left(-\frac{1}{2}, \frac{\sqrt{3}}{2}\right) = \frac{{\bf a}_2}{2} - {\bf a}_1,
\end{aligned}
\end{equation}
the hopping Hamiltonians $\mc{H}_i$ can be explicitly written as
\begin{alignat*}{3}
     \mc{H}^{}_{1}
     &=\zeta\sum_{\vec{r}}
     &&\bigg(&&\mathrm{e}^{\text{i}\phi_{1}}c_{\vec{r}}^{\dagger}c^\pdagger_{\vec{r}+\delta_{1}}+\mathrm{e}^{-\text{i}\phi_{1}}c_{\vec{r}+\delta_{2}}^{\dagger}c^\pdagger_{\vec{r}+\delta_{1}+\delta_{2}}  \\
     \nonumber
    & &&+ &&\mathrm{e}^{\text{i}\phi_{2}}c_{\vec{r}}^{\dagger}c^\pdagger_{\vec{r}+\delta_{2}}-\mathrm{e}^{-\text{i}\phi_{2}}c_{\vec{r}+\delta_{2}}^{\dagger}c^\pdagger_{\vec{r}+2\delta_{2}} \\
    \nonumber
     & &&+&&c_{\vec{r}}^{\dagger}c^\pdagger_{\vec{r}+\delta_{3}}+c_{\vec{r}+\delta_{2}}^{\dagger}c^\pdagger_{\vec{r}+\delta_{3}+\delta_{2}}+\mathrm{h.c.}\bigg),\\
\mc{H}^{}_{2}
\nonumber
&=\lambda\sum_{\vec{r}} &&\bigg(&&\mathrm{e}^{\text{i}\varphi_{1}}c_{\vec{r}}^{\dagger}c^\pdagger_{\vec{r}+\delta_{1}+\delta_{2}}+\mathrm{e}^{-\text{i}\varphi_{1} }c_{\vec{r}+\delta_{2}}^{\dagger}c^\pdagger_{\vec{r}+\delta_{1}+2\delta_{2}}  \\
\nonumber
& &&+&&\mathrm{e}^{\text{i}\varphi_{2}}c_{\vec{r}}^{\dagger}c^\pdagger_{\vec{r}+\delta_{2}+\delta_{3}}+\mathrm{e}^{-\text{i}\varphi_{2} }c_{\vec{r}+\delta_{2}}^{\dagger}c^\pdagger_{\vec{r}+2\delta_{2}+\delta_{3}} \\
\nonumber
& &&+&&\mathrm{e}^{\text{i}\varphi_{3}}c_{\vec{r}}^{\dagger}c^\pdagger_{\vec{r}+\delta_{3}-\delta_{1}}-\mathrm{e}^{-\text{i}\varphi_{3}}c_{\vec{r}+\delta_{2}}^{\dagger}c^\pdagger_{\vec{r}+\delta_{2}+\delta_{3}-\delta_{1}}+\mathrm{h.c.}\bigg),\\
\mc{H}^{}_{3}
&=\rho\sum_{\vec{r}}&&\bigg(&&\mathrm{e}^{\text{i}\gamma_{1}}c_{\vec{r}}^{\dagger}c^\pdagger_{\vec{r}+2\delta_{1}}-\mathrm{e}^{-\text{i}\gamma_{1}}c_{\vec{r}+\delta_{2}}^{\dagger}c^\pdagger_{\vec{r}+\delta_{2}+2\delta_{1}}  \\
\nonumber& &&+ &&\mathrm{e}^{\text{i}\gamma_{2}}c_{\vec{r}}^{\dagger}c^\pdagger_{\vec{r}+2\delta_{2}}-\mathrm{e}^{-\text{i}\gamma_{2} }c_{\vec{r}+\delta_{2}}^{\dagger}c^\pdagger_{\vec{r}+3\delta_{2}} \\
\nonumber& &&+ &&\mathrm{e}^{\text{i}\gamma_{3}}c_{\vec{r}}^{\dagger}c^\pdagger_{\vec{r}+2\delta_{3}}-\mathrm{e}^{-\text{i}\gamma_{3}}c_{\vec{r}+\delta_{2}}^{\dagger}c^\pdagger_{\vec{r}+\delta_{2}+2\delta_{3}}+\mathrm{h.c.} \bigg).
\end{alignat*}

While we set the NN-hopping strength $\zeta$ to unity without loss of generality, the real-valued amplitudes $\lambda$ and $\rho$ are still allowed to vary freely. To ensure compatibility with the DMRG results of Ref.~\cite{gong2019chiral}, we choose $\lambda$\,$=$\,$1.0$, and $\rho$\,$=$\,$3.0$, with the accompanying phases $\phi_{1}$\,$=$\,$\phi_{2}$\,$=$\,$\pi/2$, $\varphi_{1}$\,$=$\,$\varphi_{2}$\,$=$\,$\varphi_{3}$\,$=$\,$0$, and $\gamma_{1}$\,$=$\,$\gamma_{2}$\,$=$\,$\gamma_{3}$\,$=$\,$\pi/2$.

\begin{figure}[tb]
\includegraphics[width=\linewidth]{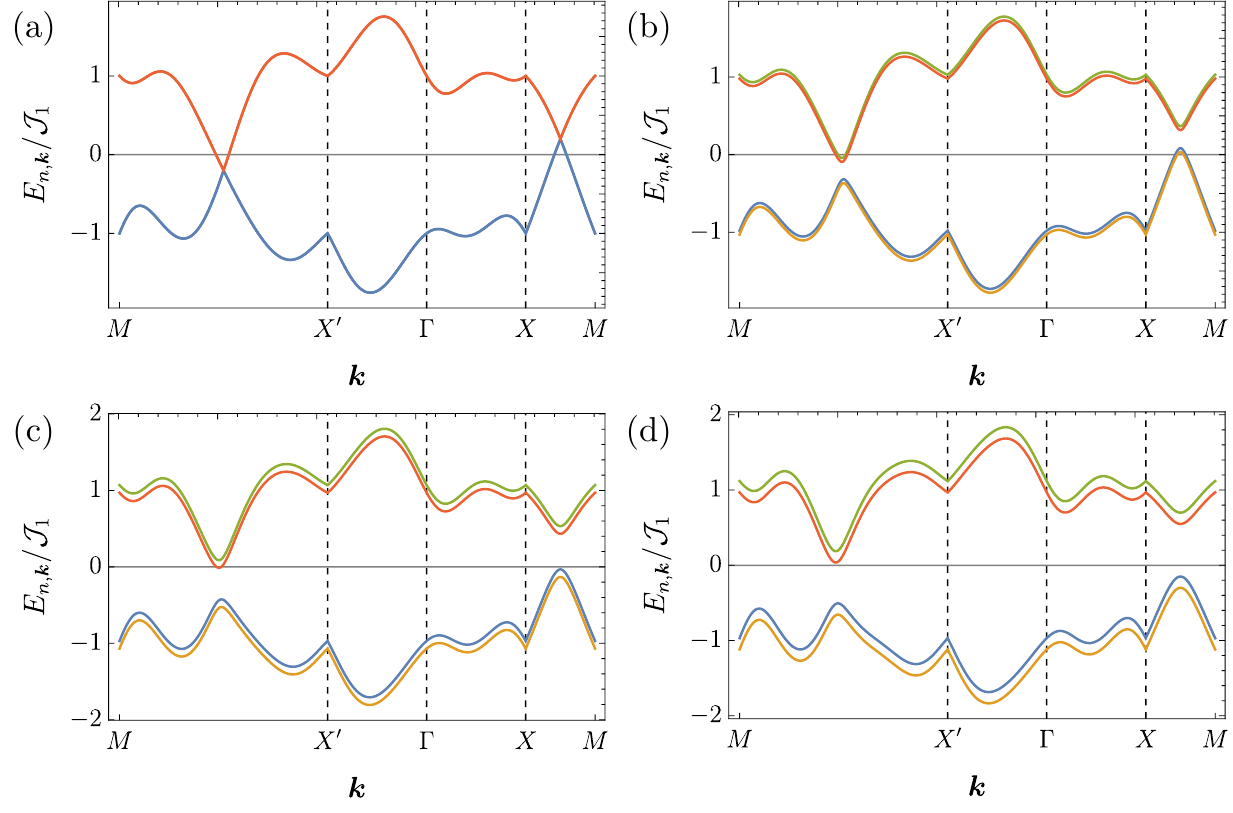}
\caption{\label{fig:ansatz}Dispersion of the mean-field Hamiltonian \eqref{eq:hH-MF2} with the same hopping amplitudes and phase factors as in Fig.~\ref{fig:h-bands} for $\mc{H}_1$ and $\mc{H}_3$. The main difference comes from the term $\widetilde{\mc{H}}_2$, in which we set $\theta_1$\,$=$\,$\pi/2$, $\theta_2$\,$=$\,$0$. The remaining parameters $(\widetilde{\mc{J}}_2, \mc{J}_3\rho, B_z)$ are taken to be (a) $(0, 0.1, 0)$, (b) $(0.1, 0.1, 0.05)$, (c) $(0.2, 0.1, 0.1)$, and (d) $(0.3, 0.1, 0.15)$. Since the modified second-NN hopping in Eq.~\eqref{eq:2NN} is assumed to originate from the orbital coupling of a magnetic field, the coupling $\widetilde{\mc{J}}_2$ is varied proportionally to the Zeeman field $B_z$.}
\end{figure}

Since $\mc{H}_2$ and $\mc{H}_3$ already break symmetries such as time reversal and reflection, we are motivated to incorporate the effect of the orbital magnetic field coupling into those terms. However, it turns out that this will not be possible for $\mc{H}_3$, which breaks the degeneracy of the Dirac points. To show that a Fermi surface cannot arise from induced field couplings in this particular model, we first recognize that in order to have a Fermi surface, the degeneracy of the Dirac points at $\vec{k}$\,$=$\,$(0,\pi/2\sqrt{3})$ and 
$(\pi,\pi/2\sqrt{3})$ must be broken. After a Fourier transformation, a general coupling contributes a  term to the Hamiltonian in the form of 
\begin{alignat}{1}
\sum_{\vec{k}}\eta_{\vec{k}}^\dag\left(v_0(\vec{k})\mathbb{I}+\sum_{i=1}^3 v_i(\mathbf{k})\sigma^i
\right)\eta_{\vec{k}},\quad \eta_{\vec{k}}=\left(
	c_{\vec{k},+},
c_{\vec{k},-}
\right)^T,
\end{alignat} 
where we have denoted $\mathbb{I}$ to be the identity matrix and $+,-$ to be sublattice labels. Under this coupling, the energy of the (possibly gapped) Dirac points are shifted by $v_0
(\vec{k})-\sqrt{v_i(\vec{k})^2}$. Note that $v_0(\vec{k})$ and $v_i(\vec{k})$ are made up of trigonometric terms with the general form \begin{alignat}{1}
\sin\left(\sum_{j=1}^3 n_j\delta_j\cdot\vec{k}+\vartheta\right),
\end{alignat} for $n_i\in\mathbb{Z}$. Now, due to the constraints from the projective translation symmetry, any allowed $v_i(\vec{k})$ term will end up not breaking the degeneracy of the Dirac points. Therefore, the condition that the two Dirac points are shifted differently means that we must have a $v_0(\vec{k})$ term that does so. Such terms come from couplings between the same sublattice ($+$ or $-$) and can take the form
\begin{alignat}{1}
\mathrm{e}^{i \phi}c_{\vec{r}}^\dagger c^\pdagger_{\vec{r}+n_1\delta_1+2n_2\delta_2}+\mathrm{e}^{-i \phi}c_{\vec{r}+\delta_2}^\dagger c^\pdagger_{\vec{r}+\delta_2+n_1\delta_1+2n_2\delta_2}
\end{alignat} 
with $\phi$\,$\ne$\,$\pm {\pi}/{2}$ for $n_1$ odd or 
\begin{alignat}{1}
\mathrm{e}^{i \phi}c_{\vec{r}}^\dagger c^\pdagger_{\vec{r}+n_1\delta_1+2n_2\delta_2}-\mathrm{e}^{-i \phi}c_{\vec{r}+\delta_2}^\dagger c^\pdagger_{\vec{r}+\delta_2+n_1\delta_1+2n_2\delta_2}
\end{alignat} 
with $\phi$\,$\ne$\,$0$ or $\pi$ for $n_1$ even. It can be checked that neither of the terms above have the correct symmetries of the magnetic field, which breaks time-reversal and reflections, but preserves their composition. Thus, field-induced couplings in this ansatz will not lead to a Fermi surface.

The orbital coupling to a magnetic field can be incorporated by modifying the second-NN term $\mc{H}_2$, which is then replaced by
\begin{alignat}{3}
\label{eq:2NN}
&\widetilde{\mc{H}}^{}_2=\sum_{\vec{r}}&&\bigg(&&\mathrm{e}^{i\theta_1}c_{\vec{r}}^\dagger c^\pdagger_{\vec{r}+\delta_1+\delta_2}+\mathrm{e}^{-i\theta_1}c_{\vec{r}+\delta_2}^\dagger c^\pdagger_{\vec{r}+\delta_1+2\delta_2}\\
\nonumber& &&+&&\mathrm{e}^{i \theta_2}c_{\vec{r}}^\dagger c^\pdagger_{\vec{r}+\delta_2-2\delta_1}-\mathrm{e}^{-i \theta_2}c_{\vec{r}+\delta_2}^\dagger c^\pdagger_{\vec{r}+2\delta_2-2\delta_1}+\mathrm{h.c.}\bigg).
\end{alignat} We have set $\theta_1$\,$=$\,$\pi/2$, $\theta_2$\,$=$\,$0$. This modified second-NN term opens up a direct gap at each Dirac point, leading to topologically nontrivial bands. We have retained the original $\mc{H}_3$, which generates a Fermi surface around each Dirac point.
The full Hamiltonian in this case, including the effect of a Zeeman coupling, is listed in Eq.~\eqref{eq:hH-MF2}.

\bibliography{TH_Refs.bib}

\end{document}